\documentclass[aps,physrev,reprint,superscriptaddress]{revtex4-2}

\usepackage{graphicx}
\usepackage{dcolumn}

\usepackage{amsmath}

\begin{document}

\title{Scanning Tunneling Spectroscopy of Subsurface Ag and Ge Impurities in Copper}

\author{Thomas Kotzott}
\affiliation{IV. Physikalisches Institut -- Solids and Nanostructures, Georg-August-Universit\"at G\"ottingen, 37077 G\"ottingen, Germany}

\author{Mohammed Bouhassoune}
\affiliation{Peter Gr\"unberg Institut and Institute for Advanced Simulation, Forschungszentrum J\"ulich and JARA, 52425 J\"ulich, Germany}
\affiliation{D\'epartement de Physique, FPS, Cadi Ayyad University, Marrakech, Morocco}

\author{Henning Pr\"user}
\affiliation{IV. Physikalisches Institut -- Solids and Nanostructures, Georg-August-Universit\"at G\"ottingen, 37077 G\"ottingen, Germany}

\author{Alexander Weismann}
\affiliation{IV. Physikalisches Institut -- Solids and Nanostructures, Georg-August-Universit\"at G\"ottingen, 37077 G\"ottingen, Germany}
\affiliation{Institut f\"ur Experimentelle und Angewandte Physik, Christian-Albrechts-Universit\"at zu Kiel, 24098 Kiel, Germany}

\author{Samir Lounis}
\affiliation{Peter Gr\"unberg Institut and Institute for Advanced Simulation, Forschungszentrum J\"ulich and JARA, 52425 J\"ulich, Germany}
\affiliation{Faculty of Physics, University of Duisburg-Essen and CENIDE, 47053 Duisburg, Germany}

\author{Martin Wenderoth}
\email[]{martin.wenderoth@uni-goettingen.de}
\affiliation{IV. Physikalisches Institut -- Solids and Nanostructures, Georg-August-Universit\"at G\"ottingen, 37077 G\"ottingen, Germany}

\date{October 12, 2021}

\begin{abstract}
We investigate single Ge and Ag impurities buried below a Cu(100) surface using low temperature scanning tunneling microscopy. The interference patterns in the local density of states are surface scattering signatures of the bulk impurities, which result from 3D Friedel oscillations and the electron focusing effect. Comparing the isoelectronic $d$ scatterer Ag and the $sp$ scatterer Ge allows to distinguish contributions from impurity scattering and the host. Energy-independent effective scattering phase shifts are extracted using a plane wave tight-binding model and reveal similar values for both species. A comparison with ab-initio calculations suggests incoherent $sp$ scattering processes at the Ge impurity. As both scatterers are spectrally homogeneous, scanning tunneling spectroscopy of the interference patterns yields real-space signatures of the bulk electronic structure. We find a kink around zero bias for both species that we assign to a renormalization of the band structure due to many-body effects, which can be described with a Debye self-energy and a surprisingly high electron-phonon coupling parameter $\lambda$. We propose that this might originate from bulk propagation in the vicinity of the surface.
\end{abstract}

\maketitle

\section{Introduction}

Scattering at impurities in a solid crucially affects material properties like electrical or thermal conductivity. Almost a century ago, the empirical Linde rule already related the residual resistivity of a dilute alloy $\rho$ with the valence difference $\Delta Z$ between host and impurity species by $\rho \propto (\Delta Z)^2$ \cite{Linde1932}. As compound in a dilute Cu alloy, e.g., Ge shows a 27 times higher residual resistivity than Ag \cite{Linde1932,Blatt1957a,Hellwege1983}. Complementary to macroscopic transport measurements, the scanning tunneling microscope (STM) yields access to the atomic scale that allows to investigate the scattering properties of single impurity atoms. This was first performed in the early days of STM in 2D scattering experiments of noble metal surface states, e.g.\ on Cu(111) or Au(111) \cite{Crommie1993,Avouris1994,Hasegawa1993}. The resulting Friedel oscillations contain information about both the defect scattering properties as well as the host. For example, on the one hand, the STM characterized a surface magnetic impurity with its Kondo signature and its scattering phase shift in the scattering pattern of a metal surface state \cite{Manoharan2000,Fiete2001,Schneider2002,Knorr2002,Quaas2004,Lounis2006}. On the other hand, by measuring the energy-dependent wave length of the scattering pattern at step edges, the dispersions of the surface states of Cu(111) and Au(111) were observed in real space \cite{Crommie1993,Hasegawa1993}. The Fourier analysis of standing wave patterns, termed as quasiparticle interference (QPI), characterizes a materials' surface 2D band structure as well as available scattering channels \cite{Sprunger1997,Petersen1998a,Hoffman2002,Pascual2004}. In specific sample systems, 2D QPI can be used to study parts of the 3D band structure \cite{Petersen1998,Marques2021}.

Of peculiar interest are properties exceeding a single electron dispersion that describe many-body effects, e.g.\ electronic interactions with phonons where self-energy corrections cause a characteristic kink around the Fermi level. Especially metal surface states as prototypical 2D system were investigated in detail \cite{Hellsing2002,Echenique2004,Hofmann2009}, e.g. by angle-resolved photoemission spectroscopy (ARPES) \cite{McDougall1995,Valla1999,Reinert2004,Shi2004}. By means of STM, the life times \cite{Burgi1999,Kliewer2000} as well as the electron-phonon coupling parameter $\lambda$ and the self-energy $\Sigma$ \cite{Grothe2013} of surface states were determined. QPI also gave access to other many-body effects including coupling to surface plasmons \cite{Sessi2015} and different phonon modes \cite{Dahm2014,Arguello2015,Choi2017}. As another approach by STM, Landau level spectroscopy offered access to 2D sample systems and their renormalized bands due to electron-phonon coupling \cite{Li2009,Zeljkovic2015}. 
 
In 3D, first, subsurface properties in metals were explored through quantum well states \cite{Altfeder1998,Altfeder2002,Kubby1992}. Then electron focusing was found to be a powerful tool to access bulk properties with an STM in real space on the atomic scale. Due to the anisotropy of a metal's band structure, the electrons propagate directionally through the bulk and form interference patterns at the surface. This was found at single magnetic impurities characterizing the host's Fermi surface \cite{Weismann2009}, at noble gas cavities \cite{Kurnosikov2008,Kurnosikov2009,Schmid1996,Schmid2000} and point-like subsurface defects \cite{Sprodowski2010}. The electron focusing effect was theoretically described \cite{Lounis2011} and applied at impurities in different contexts for the spectral signature of the Kondo effect at single atoms \cite{Pruser2011,Pruser2012,Pruser2014}, for the two-band superconducting gap of Pb \cite{Ruby2015}, and for quantum well states within layered systems \cite{Bouhassoune2014}. 

In this study, we map the surface signatures of non-magnetic impurity atoms buried in Cu(100). The scattering patterns of the isoelectronic $d$ scatterer Ag and the $sp$ scatterer Ge are compared in order to distinguish contributions from the impurity and the host. We extract similar energy-independent effective scattering phase shifts for both impurity species. Scanning tunneling spectroscopy (STS) reveals the energy-dependence of the electron focusing patterns, which act as real-space signature of the isoenergy surfaces of the host's 3D band structure due to the spectral homogeneity of the scattering phase. Most of the detected patterns is reproduced by electronic structure simulations, except from spectral anomalies intriguingly located at the Fermi energy. These can be described as the signature of a Debye self-energy that is discussed as bulk electron-phonon coupling in the vicinity of the surface.

\section{Experimental Results}
\subsection{Topographic signature of subsurface impurities}
The experiments were performed with home-built STMs operating at a temperature of $T \approx 6\,\text{K}$ and a base pressure of $p < 5 \times 10^{-11}\,\text{mbar}$. We investigate dilute Cu surface alloys with $<1\%$ of Ge or Ag impurities, respectively. In a first step of sample preparation, Cu(100) single crystals are cleaned by cycles of Ar$^+$ sputtering and annealing. Subsequently, Cu and the impurity material are simultaneously deposited from two electron beam evaporators. The sample preparation is described in detail in Supplementary Note S1. 

\begin{figure*}
	\includegraphics[width=17.2cm]{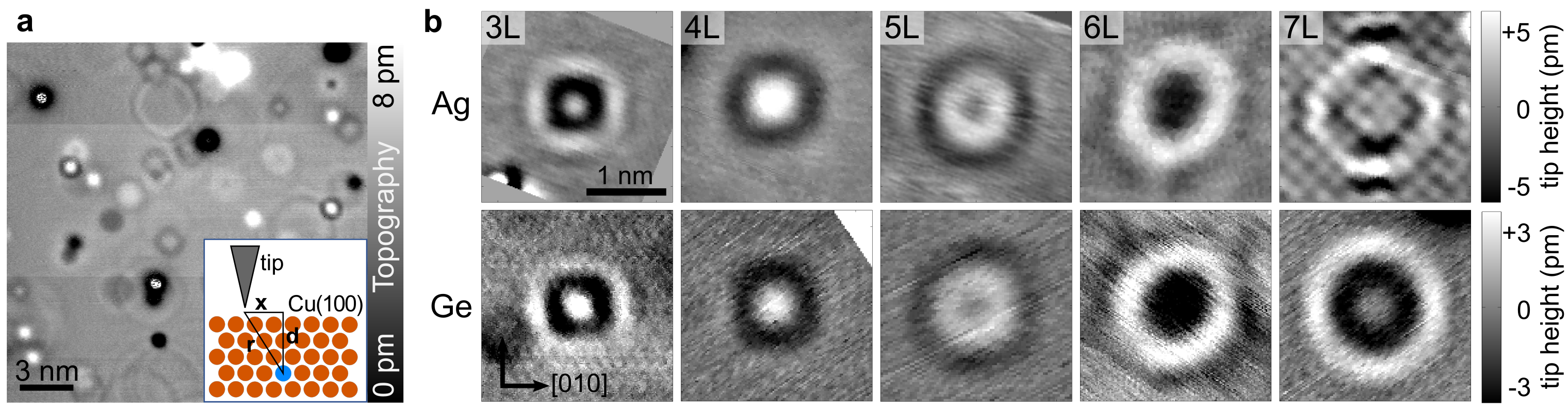}
	\caption{\label{fig:AgGeExpComp} Topographic surface signatures of subsurface non-magnetic impurities. (a) Topography ($U = -100$\,mV / $I = 0.5$\,nA) with various LDOS patterns, each one corresponding to surface defects (black spots) or buried Ge impurities (shallow ring-like features). Inset: Sketch of sample system with impurity atom in 4th layer depicted in blue. (b) Comparison of experimental surface patterns (2.4$\times$2.4\,nm$^2$) for Ag and Ge impurities for various depths. Both species show similar characteristic shapes for every layer increasing in size. The deviations for Ag 7L originate from the atomic resolution superimposed with the bulk focusing signal. Set points Ag ($U=-10$\,mV / $I$(3L,7L)=2\,nA, $I$(4L,5L)=3\,nA, $I$(6L)=0.8\,nA) and Ge (3L: $U$=$-25$\,mV/$I$=0.5\,nA; 4L,7L: $U$=100\,mV/$I$=0.4\,nA; 5L: $U$=50\,mV/$I$=0.3\,nA; 6L (plotted $z\times$0.5): $U$=100\,mV/$I$=2.0\,nA).}
\end{figure*}

Both species of non-magnetic impurities feature picometer-high Friedel oscillations as surface signature due to electron focusing. A large scale topography in Fig.~\ref{fig:AgGeExpComp}a shows interference patterns of Ge impurities at the Cu(100) surface. Each pattern corresponds to a single subsurface impurity causing a standing wave pattern of scattered electrons that propagate to the surface. The surface signatures have four-fold symmetry as defined by the crystal surface orientation and the anisotropic Cu band structure. The latter determines a specific angle under which the electrons propagate from the impurity towards the surface. Thus, a laterally more extended surface pattern corresponds to an impurity located deeper below the surface. Strong bright contrast are surface adatoms while dark contrast is assigned to surface layer impurities (termed '1st layer impurities' hereafter). The topographic contrast in the local density of states (LDOS) for subsurface impurities is with a few picometers a small fraction of a monoatomic step. (For more information see Supplementary Note S1).

In Fig.~\ref{fig:AgGeExpComp}b, we present the surface signatures of Ag and Ge impurities from 3rd to 7th monolayer (ML) in Cu(100). The topographic contrast of a single impurity can be identified and assigned a depth using atomic resolution data (c.f. Supplementary Note S2). With increasing depth, the picometer-high corrugations increase in lateral size. Each four-fold pattern has its characteristic shape of maxima and minima due to constructive and destructive interference. Comparing both species, we find the topographic signatures to be very similar in both shape and amplitude. We can reproduce the experimental data with a tight-binding plane wave model adapted from Ref.\ \cite{Weismann2009} which uses an effective scattering phase shift $\eta$ to characterize the shape of the interference pattern (for detailed information, see Appendix \ref{app:TBmodel}). The average best-fit effective scattering phase shifts of  $\eta_{\text{Ag}} = (1.32 \pm 0.04)\pi$ and $\eta_{\text{Ge}} = (1.23 \pm 0.05)\pi $ match the similarity of both species as observed in topography data.

\subsection{Spectroscopic signature of subsurface impurities}
We resolve the energy-dependence of the electron focusing patterns by using scanning tunneling spectroscopy. For each voltage, STS probes the interference pattern of the subset of states with wave vectors of the corresponding energy. Data was processed with topography normalization to obtain spectroscopic information for a constant height contour. For analyzing the change in LDOS due to the impurity, we subtract averaged spectra of the pristine Cu surface far away from the buried atom from the differential conductance around the impurity. As the topographic patterns only show a height contrast of a few picometers, the remaining spectroscopic signal shown in the following corresponds to approximately 1\% of LDOS modulation obtained in the raw data. 

\begin{figure}
	\includegraphics[width=8.6cm]{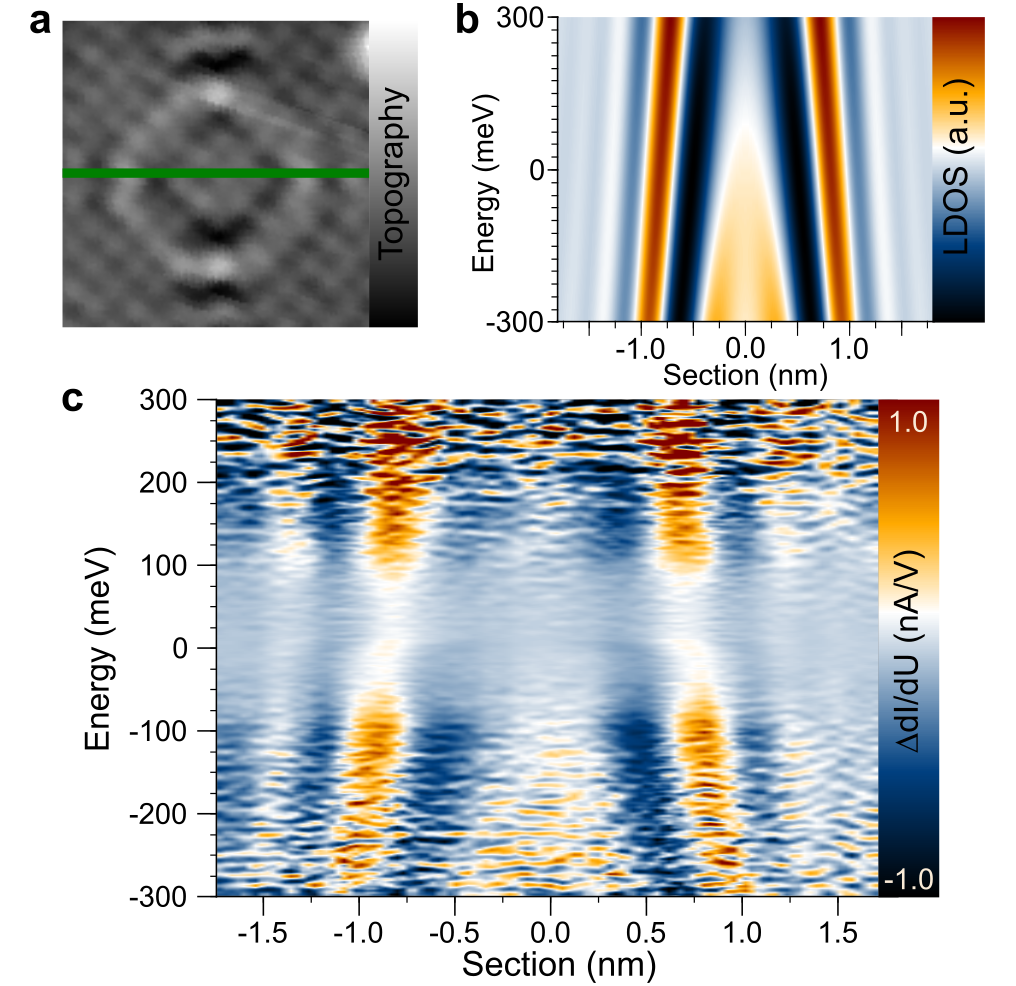}
	\caption{\label{fig:Ag7L-STS} Spectroscopic signature of 7th ML Ag impurity in Cu(100). (a) Topography ($U=-10$\,mV/ $I=2$\,nA, 2.4$\times$2.4\,nm$^2$) shows four-fold symmetry and beam-like continuations along crystal axes superimposed with atomic resolution. Crystal [010] direction is indicated by green line. (b) Spectroscopic section along [010] direction, simulated by the tight-binding model ($h$=5\,\AA, $\eta$=1.0$\pi$). (c) Experimental spectroscopic section along [010] direction. The difference $\Delta$d$I$/d$U$ with respect to the clean Cu surface is color-coded. The maxima and minima of the interference pattern match quantitatively with the simulation in (b). At the Fermi level, an additional bending of the pattern is found.}
\end{figure}

We find the spectroscopic signatures of buried impurities to be very similar for both species, Ag and Ge, which agrees with the findings for the topography data. In Fig.~\ref{fig:Ag7L-STS} we show a 7th ML Ag impurity along with its spectroscopic signature. In the topography, the four-fold symmetry is well-resolved. A spectroscopic line section through the pattern along the crystal's [010] axis is presented in Fig.~\ref{fig:Ag7L-STS}c. The color-code depicts the difference in differential conductance $\Delta$d$I$/d$U$ with respect to the pristine Cu surface. The main characteristic is an inwards movement of the positions of maxima and minima due to energy dispersion. A higher electron energy leads to a shorter wave length which fulfills the same interference condition for shorter distances. Dispersion is also the dominating effect in simulated spectroscopy data, calculated using the tight-binding model (Fig.~\ref{fig:Ag7L-STS}b). The general shape of the simulation, which is given by the positions of maxima and minima lines, matches well with the experiment (Fig.~\ref{fig:Ag7L-STS}c). 

\begin{figure}
	\includegraphics[width=8.6cm]{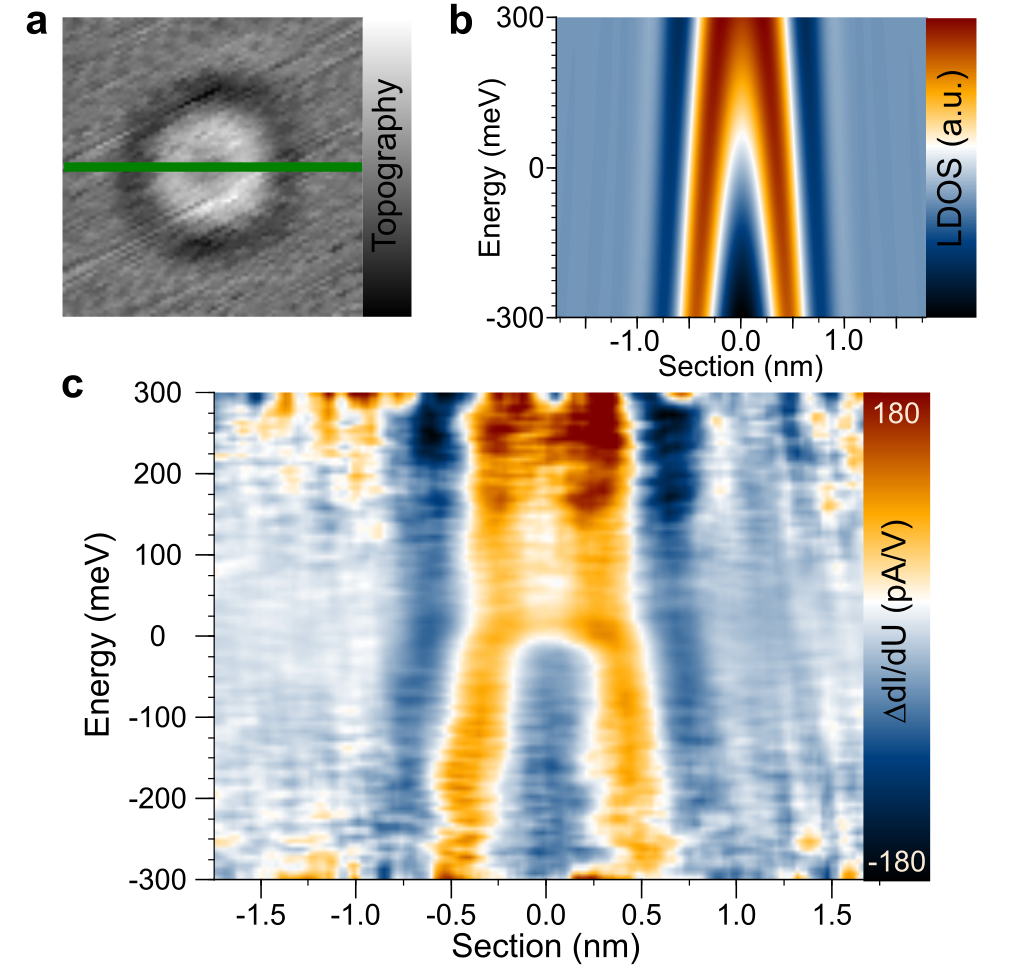}
	\caption{\label{fig:Ge5L-STS} Spectroscopic signature of 5th ML Ge impurity in Cu(100). (a) Topography ($U = -$50\,mV / $I = $0.3\,nA, $2.6\times2.6$\,nm$^2$) with four-fold symmetric surface pattern. Crystal [010] direction is indicated by green line. (b) Spectroscopic section along [010] direction, simulated by the tight-binding model ($h$=7\,\AA, $\eta$=1.25$\pi$). (c) Experimental spectroscopic section along [010] direction. The difference $\Delta$d$I$/d$U$ with respect to the clean Cu surface color-coded. The maxima and minima of the interference pattern match quantitatively with the simulation in (b). At the Fermi level, an additional bending of the pattern can be observed.}
\end{figure}

A spectroscopy data set of a 5th ML Ge impurity is shown in Fig.~\ref{fig:Ge5L-STS} with topography, spectral section and tight-binding simulation. As for Ag, we find the spectroscopic signature of a Ge impurity to be governed by electron dispersion with good agreement between experimental and calculated patterns. A single effective scattering phase can describe the data in the whole energy range without any strong additional resonances at element-specific energies. As for the topographies, the spectroscopy data for Ag and Ge is very similar for the same impurity depths.

An additional spectroscopic feature is found for both species at the Fermi level. In Fig.~\ref{fig:Ag7L-STS}c and Fig.~\ref{fig:Ge5L-STS}c, a bending of the interference pattern can be seen around 0 eV bias voltage where the pattern shifts towards the center in an immediate jump as if it experiencing a small phase shift. Considering a larger energy interval, the dispersion is unaffected by this anomaly.  We observed these characteristics in multiple data sets for both species and different depths, including a Ge impurity buried in the 18th ML (c.f. Supplementary Note S4). The additional feature is not found in simulations of the tight-binding model, so it is beyond the scope of this single-particle band structure calculation.

\section{Ab-initio Calculations}
Complementary to the experiment and the simplified tight-binding model, we performed ab-initio calculations using density functional theory (DFT) as implemented in the full potential Korringa-Kohn-Rostoker Green function (FP-KKR-GF) method \cite{Papanikolaou2002} with the local density approximation (LDA) as parameterized by Vosko, Wilk and Nusair \cite{Vosko1980}. The theoretical and numerical approach is similar to that used in Refs.\ \cite{Weismann2009,Lounis2011, Pruser2014, Bouhassoune2014} and is described in more detail in Supplementary Note S3.

\begin{figure}
	\includegraphics[width=8.6cm]{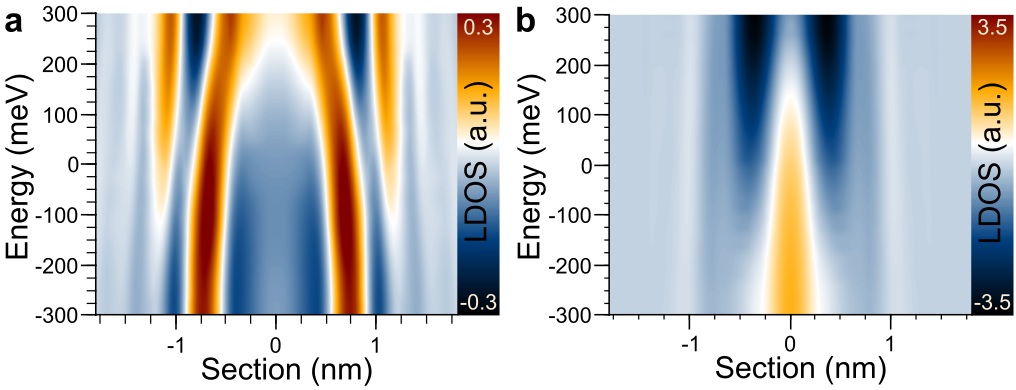}
	\caption{\label{fig:DFT-STS} Spectroscopic sections calculated by ab-initio KKR-DFT calculations. (a) 7th ML Ag impurity. The pattern matches with the corresponding experimental data. (b) 5th ML Ge impurity. The pattern is dominated by a contrast-changing feature in the center and does not resemble the experimental counterpart. In the outmost parts of the pattern, weak oscillatory components, similar to the Ag pattern, are found. The overall amplitude is a multiple of the calculated signal for Ag.}
\end{figure}

In Fig.~\ref{fig:DFT-STS}, we show surface spectral sections along the crystal symmetry axis as calculated from the KKR-DFT potentials. Figure~\ref{fig:DFT-STS}a and Fig.~\ref{fig:DFT-STS}b show a 7th ML Ag impurity and a 5th ML Ge impurity, corresponding to the experimental sections presented in Fig.~\ref{fig:Ag7L-STS}c and Fig.~\ref{fig:Ge5L-STS}c. The calculated spectra show richer patterns than the tight-binding model because the full crystal potential is considered. The pattern shape shows more detailed features than an almost linear dispersion and the LDOS amplitude modulates with energy. The total amplitude of the signal for Ge is multiple times higher than for Ag.

Comparing ab-initio calculation and experiment, the Ag data shows good qualitative agreement reproducing the overall pattern shape. With only small corrections to the ab-initio calculations' parameters (e. g. a shift of the energy scale by $150$\,mV or slightly corrected scattering phase shifts), the positions of the pattern's minima and maxima for the Fermi level match also quantitatively (c.f. Supplementary Note S3). For Ge, the patterns differ from each other. The calculation's rich features in the center including the contrast inversion from positive to negative with increasing energy are not found in the experiment, which presents a pattern that does not change qualitatively with energy and only shows a small dispersive shift. The pattern shape of the ab-initio data in the outer region ($r > 8$\,\AA) resembles better the overall experimental pattern shape and is also similar to the Ag data. This different behavior for Ag and Ge data holds also for spectroscopic sections of other depths of both species.

\section{Discussion}
\subsection{Contributions of orbital components}
Surface interference patterns are detected by STM even for atoms buried many layers below the surface despite the fact that non-magnetic impurities are weak scatterers. By the specific design of this experiment, we probe only the contributions of coherently scattered electrons for which the incoming and the scattered waves interfere constructively or destructively at the surface. The shape and size of a pattern are determined by two main ingredients: firstly, the scattering event, i.e., the scattering properties of the impurity atom and the electronic states it interacts with and, secondly, the electron propagation between surface and impurity.

In the experiment, we find the topographic signature of both impurity species to be very similar. From the literature, one would expect to find severely different scattering characteristics as the species differ strongly in both residual resistivity as well as the predicted scattering channels. In scattering theory, the scattering process at an impurity is described by its influence on host states being scattered. These electron wave functions are expanded in spherical harmonics, usually limited to $s$, $p$, and $d$ orbitals. The influence on their asymptotic behavior by the impurity potential is parametrized by orbital-dependent scattering phases which are the main contribution to the scattering matrix $T$. These considerations are essential to the DFT-KKR ab-initio calculations, where the difference of element-specific scattering matrices determines by its amplitude how strong the scattering is in the respective channel for a specific impurity in a specific host. In other words, defects scatter especially in those channels, where the impurity differs most from a host atom. In this framework, strong scattering means that the host's states of a specific orbital character are especially disturbed and it does not describe scattering during bulk state propagation in the pure host or how many states of the respective orbital character are present. For the transition metal Ag in the isoelectric Cu, scattering at a $3d$ virtual bound state gives rise to the low residual resistivity \cite{Mertig1999}. As the resonance state is located far from the Fermi energy \cite{Braspenning1982}, leading to only little interaction with the electrons at the Fermi level, a small scattering amplitude of $d$ scattering is predicted. For the $sp$ impurity Ge in Cu, the perturbations of the crystal potential due to the significant additional local valence charge \cite{Braspenning1982} of three additional electrons in $s$ and $p$ orbitals constitutes the main scattering process, leading to higher scattering rates especially for electrons of $s$ and $p$ character.

These predicted scattering intensities agree with the scattering amplitudes calculated in our ab-initio calculations (see Supplementary Note S3). While the scattering amplitude for Ge in Cu is three to four times higher for $s$ and $p$ orbital characters than for the $d$ contribution, all Ag scattering amplitudes are half or less of the smallest value for Ge. This is reflected in the calculated LDOS amplitudes (see Fig.~\ref{fig:DFT-STS}) where for Ge the signal is up to a factor of 7 higher than for Ag in the same depth. Even more strikingly, also the calculated pattern shape strongly differs between both species (c.f. Supplementary Figure S6).

While ab-initio calculations and experiment match for Ag impurities, there are large deviations for Ge. In order to explain these differences, the importance of the orbitals for the scattering process leads us to investigate if LDOS features of the simulations can be assigned to specific orbital components. For this, it has to be noted that the ab-initio calculations contain two orbital-dependent components, namely the impurity scattering as well as the propagation of Cu bulk states within the crystal lattice. For the latter, we can apply the tight-binding model (c.f. Appendix \ref{app:TBmodel}). We determine the orbital character of the host metal's Bloch states in order to analyze how different orbital contributions are distributed over the Fermi surface and how this affects the electron focusing patterns at the crystal surface.

\begin{figure*}
	\includegraphics[width=17.2cm]{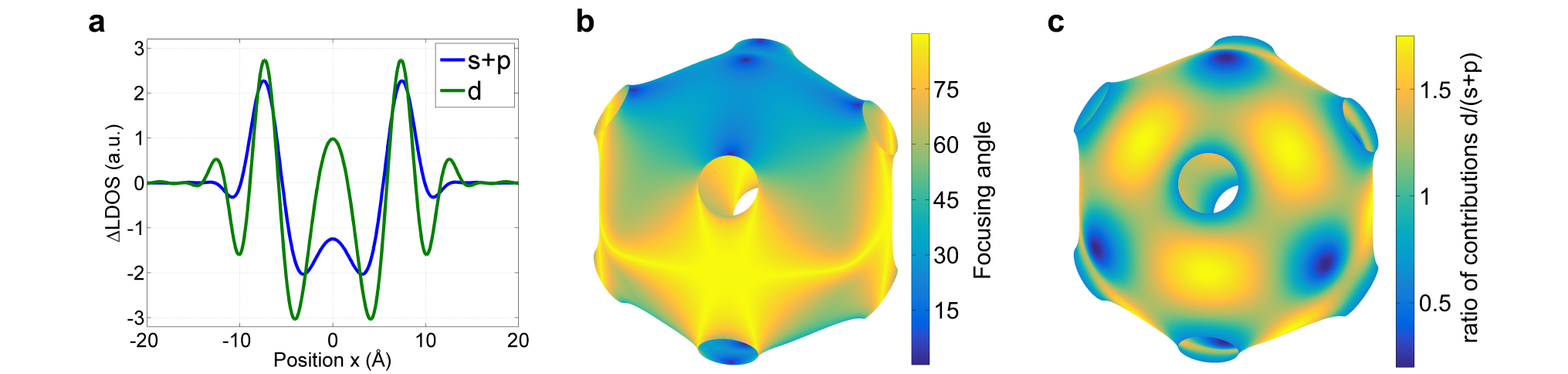}
	\caption{\label{fig:FermiSurface} Electron focusing and orbital character. (a) Orbital-resolved tight-binding simulation of the surface signature of a 7th ML Ge impurity for $U$=0\,eV with tip height $h=7$\,\AA. The outmost peak at $x=12$\,\r{A} is only visible for the $d$ contribution (green). (b) The focusing angle with respect to the surface normal for each state on the Cu Fermi surface. The strong focusing regions reveal angles of $30^\circ-40^\circ$ and $50^\circ-60^\circ$. The areas in $\langle 100 \rangle$ and $\langle 111 \rangle$ directions show angles close to $0^\circ$ for orientations towards the crystal surface. (c) Tight-binding calculation of the Cu Fermi surface including the orbital character. The ratio of $d$ contributions and ($s+p$) contributions is color-coded for each state. Regions dominated by $d$ orbital character coincide with flat regions that mainly contribute to focusing.}
\end{figure*}

In Fig.~\ref{fig:FermiSurface}a the contributions of Cu $sp$ and $d$ states in an exemplary interference pattern are shown as line sections as calculated with the tight-binding model. For each orbital, both propagation paths to and from the impurity are described by the orbital-specific propagator. For equal scattering amplitudes for $sp$ and $d$ states, the surface signal of $d$ electrons is around 60\% of the whole signal and, therefore, larger in amplitude than the combined $sp$ contributions. At the outmost angles, the maximum at 12\,\AA\ is only found in the $d$ signal. This is representative for the signatures of impurities in all depths. In the center of the pattern, contributions from all electronic orbitals superimpose, while in the most outward region, the surface pattern is strongly dominated by $d$ contributions.

In more detail, we can link this observation on a simulated topography to the host's Fermi surface: the interference patterns are constituted by bulk electrons that propagate from the impurity to the crystal surface in the direction of their group velocity. Electron focusing describes the fact that flat regions of the Fermi surface accumulate states with a certain propagation angle and, thus, cause a strong surface LDOS signal \cite{Weismann2009,Lounis2011}. Figure~\ref{fig:FermiSurface}b shows that for Cu(100) the focusing regions are especially the flat facets near the $\langle 110 \rangle$ directions (between the necks) with a focusing angle of $30^\circ$ to $40^\circ$. The regions in $\langle 100 \rangle$ direction and around the necks ($\langle 111 \rangle$) show smaller focusing angles. From the tight-binding model, we determine the orbital character of states at different regions of the Fermi surface as shown in Fig.~\ref{fig:FermiSurface}c in agreement with previous calculations \cite{Mertig1999,Mustafa2016}. For all areas contributions from all orbitals are present with varying intensity. We find that the regions dominated by $d$ character (bright) coincide with the flat focusing regions, whereas the neck and belly regions show stronger contributions from $s$ and $p$ (dark) whose smaller focusing angle points more towards the center of the interference pattern. Consequently, the orbital-resolved characteristics of the Fermi surface support what has already been visualized in the simulated orbital-resolved interference pattern.

With this knowledge, we can interpret the differences in the data of buried Ge atoms. In the ab-initio calculations, the outer, weak-contrast dispersion-like features are assigned to electrons with $d$ character. In the center area, a very dominant feature from $sp$ (because of strong $sp$ scattering at the Ge impurity) masks a weak $d$ electron background. As in the experiment such a $sp$ feature is not visible, the signal of the experimental Ge topography is attributed to $d$ electrons, which is supported by DFT in the outer regions. As our experiment is only sensitive to coherently scattered electrons interfering at the surface, we conclude that Ge shows such coherent scattering only for Cu electrons with $d$ character and attribute the missing $sp$ signal to additional incoherent processes in the sample system, which are not being considered in the ab-initio calculation.

We propose that these incoherent scattering processes occur at the impurity while we assume incoherence introduced during electron propagation to play only a minor role. Previous ab-initio studies have found strongly $k$-dependent relaxation times for bulk electrons regarding electron-phonon interactions for different regions of the Fermi surface \cite{Mustafa2016} with shorter times at necks and in $\langle 100 \rangle$ directions and longer times at flat regions which could explain stronger damping for the low angle regions than for high angles. However, as the relaxation times for room temperature range from 18\,fs to 54\,fs for different parts of the Fermi surface, the corresponding mean free paths are up to an order of magnitude larger than the propagation distance between surface and impurity in our low-temperature experiment. At low temperatures and including electron-electron interaction, bulk electron coherence has been assumed to be several dozens of nanometer even for energies up to 1\,V from the Fermi energy \cite{Campillo2000,Knoesel1998}. Therefore, the coherence lengths are too large for the $sp$ contributions to be damped by scattering within the propagation distance of only few nanometers. Instead, we suggest that the $sp$ scattering process at the Ge impurity is partially not coherent which would strongly reduce the signal in our experiment. Within these considerations, the substantially increased residual resistivity of Ge impurities in Cu with respect to Ag could be attributed to decoherent scattering of $sp$ electrons at the impurity.

Having discussed the origin of the differences between ab-initio calculations and experimental data for Ge, it remains the question why Ag and Ge show almost identical surface signatures in the experiments. For Ag, our DFT calculations and literature describe $sp$ scattering to be less dominant than $d$ scattering and a similar effective phase shift for all orbital contributions. This agrees with the simulated LDOS data that does not exhibit such a strong feature as Ge, but instead shows dispersive lines over the whole lateral range. These consist of all orbital characters and especially $d$ electrons as these are the strongest orbital contribution in the electron focusing propagation. For Ag, the DFT calculation matches with the experimental data. When comparing both species, the KKR scattering phases for Ag are similar to the Ge scattering phase for the $d$ orbital which is, in this case, the only imaged orbital channel. Hence, this can explain why the experimentally acquired topographies resemble each other for both species.

\subsection{Signatures of Many-Body Effects}
So far the analysis has shown that the surface signatures of buried impurities are also determined to a large extent by the bulk electron properties. This allows to compare scattering at two weak scatterers with different properties, here Ag and Ge impurities, and to separate the contributions from the specific impurity and general properties of the host. Thus, we use the real-space spectroscopic data to characterize the electronic structure of the Cu host around the Fermi level.

Similarly to the analogous STS studies for 2D surface states, here, we perform an experiment for standing waves of 3D Cu bulk electrons reflected at sub-surface point defects. In previous studies for single bulk impurities in Cu \cite{Pruser2012,Pruser2011}, the Kondo signatures of buried magnetic atoms were investigated. As magnetic impurities induce a phase shift around zero bias due to the Kondo resonance, an analysis of features from the host with weaker impact was not possible. 

Analysing the spectroscopic data of the non-magnetic atoms Ag and Ge in Fig.~\ref{fig:Ag7L-STS} and and Fig.~\ref{fig:Ge5L-STS}, respectively, the main observation in the range from $-300\,$mV to 300\,mV is a reduction of the interference pattern's lateral size for increasing bias voltage. This can be directly attributed to the dispersion relation of Cu bulk electrons as visualized in the tight-binding model (c.f. Figs.~\ref{fig:Ag7L-STS}b, \ref{fig:Ge5L-STS}b). The reduction of the wave length for higher energies, which leads to a smaller ring structure, is the real space mapping of increasing electron momentum for higher energy.

In addition to this expected observation, the experimental data shows significant deviations from the simple, almost-linear bulk dispersion around the Fermi level. A bending of the pattern shape around zero bias is detected for both species in an energy range of approximately $\pm30$\,mV. This energy scale is common for lattice vibrations and it is known for the dispersion in momentum space that electron-phonon interaction leads to a renormalization of electron band structure, often manifested as 'kink' in the corresponding energy interval around the Fermi level \cite{Grimvall1981}. We propose that in our experiment we map the real space signature of electronic coupling to phonon modes of the host.

In many-body theory, electron-phonon coupling is described by electron-like quasiparticles that include the interaction with phonons in their dispersion relation by self-energy corrections. These are determined by the Eliashberg function $\alpha^2F(\omega)$ or the averaged electron-phonon coupling parameter $\lambda$ \cite{Grimvall1981} whose coupling strength is given on the one hand by the available electron and phonon states, i.e. the respective densities of states, and on the other hand by the matrix element linking the two. While the real part of the self-energy shifts the center energy for a given $k$-value, the imaginary part causes a broadening in the spectral function due to finite quasiparticle lifetimes. In order to estimate the effect of such a band structure renormalization on the real space signatures, we implement the real part of the self-energy into the tight-binding model. For the Cu phonons, we use the Debye model with an Eliashberg function $\alpha^2F(\omega)=\lambda (\omega/\omega_D)^2$ for $\omega < \omega_D$ and zero elsewhere, and the Debye frequency $\hbar \omega_D = $30\,meV \cite{Gayone2005,Gross2014}. The influence of the imaginary part of the self-energy is investigated in a simple 1D toy model with linear dispersion. We find broadening especially for states with energies $E\geq\omega_D$, but no indications for strong changes in $\lambda$ or the overall pattern shape (see Appendix \ref{app:EPH1Dmodel}).

\begin{figure*}
	\includegraphics[width=17.2cm]{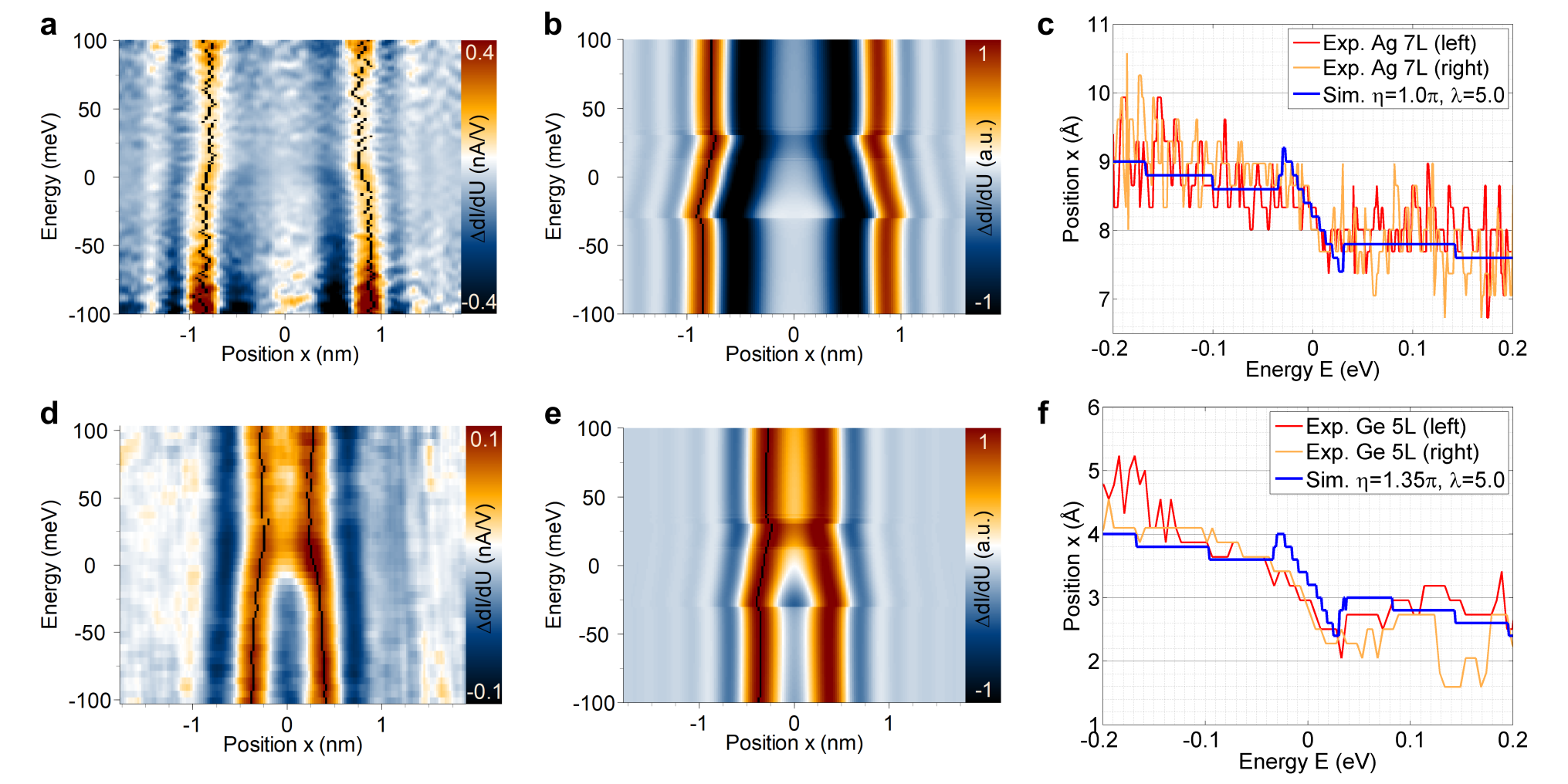}
	\caption{\label{fig:EPHAgGe} Spectroscopic signatures of electron-phonon coupling. (a,d) Experimental STS data for Ag 7ML ($U = -300$\,mV / $I = 2.0$\,nA) and Ge 5ML ($U = -300$\,mV / $I = 0.7$\,nA) shows bending of interference pattern in an energy interval around the Fermi level. The maximum of the pattern (left and right) is labelled (dark blue). (b,e) Tight-binding simulation for the respective impurities including electron-phonon coupling in the Debye model ($\lambda=5.0$, $\omega_D=30$\,meV, $h=5$\,\AA). (c,f) 1D Comparison of the maximum position in experimental (red/orange) and simulated (blue) data. While for higher energies the position is almost constant, in an interval [$-\omega_D,\omega_D$] both signals show a concise step.}
\end{figure*}

Figure \ref{fig:EPHAgGe} shows the results and a comparison to the experimental data of 7th ML Ag and 5th ML Ge impurities. The tight-binding simulations including electron-phonon coupling reveal distinctive features, especially for $|E|<\hbar\omega_D$ (Fig.~\ref{fig:EPHAgGe}b and \ref{fig:EPHAgGe}e). The patterns exhibit strong deviations from the simulations without many-body effects (Fig.~\ref{fig:Ag7L-STS}b and \ref{fig:Ge5L-STS}b) around the Fermi level, which are the real space signature of the well-known kink in momentum space. Comparing calculated and experimental data, we find good agreement for both species. For example, for the Ge pattern (Fig.~\ref{fig:EPHAgGe}d), the maxima almost join at around 25\,meV before bending outward again, which is reproduced in the calculation (Fig.~\ref{fig:EPHAgGe}e). In the simulations, discontinuities are present at $\omega_D$, which are linked to the used Debye model that introduces an abrupt cut in the phonon density of states. For a real Cu phonon spectrum, the self-energy shows a smoother course that leads to weaker features at $\omega_D$ (c.f. Supplementary Note S6). Furthermore, the simulation assumes a temperature of $T = 0$ K, while the experimental data is smoothened by thermal broadening on the scale of a few millivolts, blurring possible features at the Debye energy. The comparison in Fig.~\ref{fig:EPHAgGe}c and Fig.~\ref{fig:EPHAgGe}f shows the shift of the spatial position of the pattern's maximum with energy, as labelled in the 2D data. Good agreement is obtained with almost constant values for $|E|>$0.1\,eV and a steep step within $[-\omega_D,\omega_D]$.

To describe the experimental data, we use an electron-phonon coupling parameter of $\lambda = 5.0$, which is very high compared to the Cu bulk value $\lambda_{\text{Cu}} = 0.15$ or even materials with strong coupling like PbBi alloys ($\lambda\approx 3.0$) \cite{Grimvall1981,Grimvall1976}. A calculation similar to Fig.~\ref{fig:EPHAgGe} with the bulk value $\lambda_{\text{Cu}}$ does not reproduce the experimental data (c.f. Supplementary Note S5). Although the tight-binding model does not include all interactions and therefore is not expected to quantify exactly the experimentally found phenomena, this large deviation needs to be discussed. 

Instead of coupling with delocalized bulk phonons, the spectroscopic features could be linked to local bulk vibrational modes induced by the impurity atom at the impurity site or extended over several nanometers around the defect. However, as Ag is significantly heavier than Ge, different local vibrational modes would be expected, but both species' signatures show a similar strength of deviation from the single-particle dispersion. Furthermore, also electronic resonances, e.g.\ originating from a hybridization of bulk states with the impurity, which generally cause a phase shift, would very unlikely reveal a resonance exactly at the Fermi level for both species.

In the literature, the Cu bulk coupling parameter $\lambda$ shows a directional variance of a factor of 2-3, leading to values as $\lambda(k)=0.08...0.23$ \cite{Lee1970,Khan1982,Mustafa2016}. Although our model uses the simplifications of no $k$-dependence and particle-hole symmetry in the self-energy, this alone cannot account for a factor of 30 between $\lambda=5.0$ and the bulk value. According to ab-initio calculations, the bulk Eliashberg function $\alpha^2F(\omega)$ generally follows the phonon density of states without unusually large coupling of particular parts of the phonon spectra to electronic states \cite{Bose2008}. Yet, our experiment is very specific about electron $k$ values, as every point $(x,y)$ probes only a specific region of the isoenergy contour with the corresponding angle of group velocity, so the general statement might not hold for an electron focusing signal.

We propose that the vicinity of the surface enhances the electron-phonon coupling parameter for our experiment. Missing translational symmetry breaks the momentum conservation perpendicular to the surface which opens up additional scattering processes with more available (bulk and surface) phonon modes as well as more available final electronic states.

Electron-phonon coupling of metal surface states has extensively been studied by surface-sensitive techniques \cite{Echenique2004,Hofmann2009}. For various systems, Mo(110) \cite{Valla1999}, Cu(111) \cite{McDougall1995,Reinert2001,Reinert2004,Tamai2013}, Cu(110) \cite{Jiang2014} as well as the Pb(110) bulk state \cite{Reinert2004}, values of $\lambda$ similar to the respective $\lambda_{\text{bulk}}$ were obtained by ARPES \cite{Grimvall1981}. In STM measurements, many-body effects due to electron-phonon coupling of the Ag(111) surface state and other 2D systems were quantified by the coupling parameter $\lambda$ and the self-energy which were found to match the literature values from ARPES and the bulk \cite{Grothe2013,Li2009,Zeljkovic2015}.

Further studies have shown that electron-phonon coupling can be enhanced compared to an average parameter $\lambda$, because the coupling of specific electronic and phonon modes can show strong variations as well as an energy-dependence \cite{Plummer2003,Sklyadneva2009}. Different interactions within the electronic surface band, with bulk electrons as well as with bulk phonon states and surface modes contribute unequally to the lifetime of surface states \cite{Li1998,Kliewer2000,Burgi1999}, e.g. for Cu(111) with strong coupling to the surface Rayleigh mode and interaction with specific bulk phonons \cite{Eiguren2002,Tamai2013}. By extracting the Eliashberg function an enhanced $\lambda$ was obtained for the $\mathrm{Be}(10\overline{1}0)$ surface due to coupling to low energy surface modes \cite{Shi2004}. These findings were crucial to understand the very high value of $\lambda \approx 1$ for the Be(0001) surface \cite{Sprunger1997,Balasubramanian1998,Hengsberger1999,LaShell2000,Li2019,Osterhage2021}, four times higher than the bulk value \cite{Grimvall1981}. Strong variations of the electron-phonon coupling parameter are also discussed for quantum well systems, e.g.\ thin Ag films \cite{Valla2000,Kralj2001,Luh2002,Paggel2004,Mathias2006}.

It becomes clear that for a full picture of electron-phonon interaction, all available final electronic states, available phonon modes, and the matrix elements for the scattering processes have to be taken into account at a state-specific level. This can play a key role in our experiment because we investigate a complex 3D system, which is dominated by the bulk band structure and additionally influenced by the vicinity of the surface and the atomic defects. Although the origin of the high electron-phonon coupling value of $\lambda \approx 5$ remains unclear, such an enhanced phase space of possible interactions might be a reason why the simulations including a self-energy from many-body effects can describe the experimentally observed features.

\section{Conclusions}

In summary, we perform a 3D scattering experiment at single non-magnetic impurities in a metal, detecting surface signatures of impurities buried up to 17 monolayers below the Cu(100) surface. The interference patterns at the surface characterize both impurity scattering properties of Ag and Ge as well as the host electronic structure of Cu. Analysing the topographic patterns with a plane wave tight-binding model, we extract similar effective scattering phase shifts for both impurity species. Following a comparison with ab-initio calculations, we propose that incoherent $sp$ scattering at Ge impurities is crucial for the main scattering channel of this dilute alloy. As a result, the interference patterns of Ge impurities resemble those of Ag.

Due to the spectral homogeneity of impurity scattering, the surface signatures act as real space probe of the host's electronic dispersion, which we are able to map in good agreement with electronic structure calculations. We resolve small corrections in the form of a distinct bending of the interference pattern around the Fermi level that can be described with a Debye self-energy. It visualizes in real space the quasiparticle band structure of renormalized Cu bulk bands including many-body interactions. The high value of the electron-phonon coupling parameter $\lambda$ could result from state-specific interactions and the vicinity of the surface in the investigated bulk system. 

The used experimental approach can be versatilely applied to other systems to investigate scattering properties of subsurface impurity atoms as well as electronic properties of the host material. While for magnetic impurities the Kondo signature masks the host's features at the Fermi level, non-magnetic impurities or inherent defects can be functionalized to probe the electronic structure of both bulk and surface nanostructures even close to zero bias. For a simple material such as a noble metal, we unveil non-trivial features that we propose to be emerging from phonons coupling to electrons. As the miniaturization of nanoelectronic devices continues, the presented approach can be a useful tool to understand fundamental processes on the atomic scale in order to identify electronic interactions.

\appendix

\section{Description of surface signatures with tight-binding model \label{app:TBmodel}}
\begin{figure}
	\includegraphics[width=8.6cm]{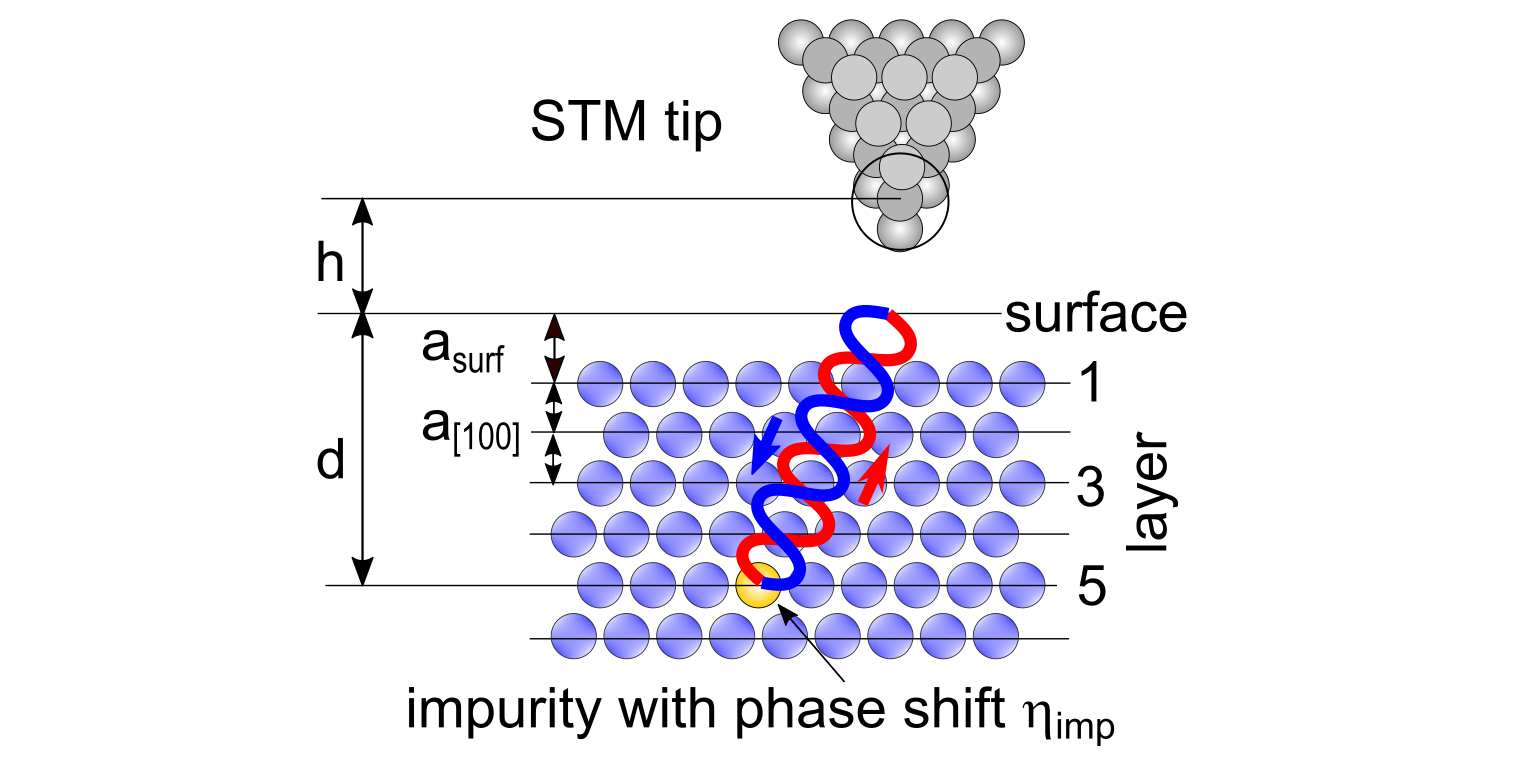}
	\caption{\label{fig-app:SketchExpPW}Sketch of the tight-binding plane wave model. Surface interference patterns result from interference between incoming (blue) and outgoing (red) electron waves, which oscillate within the crystal and experience a scattering phase shift $\eta_{\text{imp}}$ at the impurity (yellow). The tip probes the surface signature with a distance-dependent exponential decay. This distance is labeled with $h$. The depth of the impurity is given by $d$. The surface layer is referred to as the 'first layer'.}
\end{figure}

The surface signatures are analyzed and reproduced with a tight-binding plane wave (TB) model adapted from Ref.~\cite{Weismann2009}. It describes the interference of an outgoing electron wave and an incoming wave scattered at the impurity (see Fig.~\ref{fig-app:SketchExpPW}). The Bloch states are assumed to be plane waves without considering the lattice periodic part, but including the dispersion of the Cu bulk band structure, as calculated in a tight-binding calculation \cite{Papaconstantopoulos1986,Weismann2008}. For simplicity, the scattering event is reduced to an energy-independent, orbital-independent effective scattering phase shift $\eta_{\text{imp}}$ that the wavefunction collects at the impurity and a corresponding unitary scattering matrix $T=\exp(i \cdot \eta_{\text{imp}})$. The change in LDOS with respect to the pure Cu surface is given with the Green's function $G_0$ as propagator and $\mathbf{x}_{\text{imp}}$ as the position by
\begin{eqnarray}
 	\Delta\text{LDOS}(\mathbf{x},\epsilon) =& \nonumber\\
	 -\frac{1}{\pi} \operatorname{Im} [G_0(\mathbf{x},&\mathbf{x}_{\text{imp}},\epsilon)
	T(\mathbf{x}_{\text{imp}},\epsilon)
	G_0(\mathbf{x}_{\text{imp}},\mathbf{x},\epsilon)]~~
	\label{eq:TB}
\end{eqnarray}
The LDOS($\mathbf{x}$,$\epsilon$) at the surface exponentially decays into the vacuum towards the STM tip. 

We use the effective scattering phase shift $\eta_{\text{imp}}$ of the TB model to describe and quantify the shape of the experimental surface signatures. It can take values between 0 and 2$\pi$ and is assumed to be constant for different bulk impurity depths. This is reasonable as for bulk conditions the surroundings of an impurity are always identical, independent of location within the single crystal. The other free parameters of the model are the height of the tip $h$ above the sample, the distance $a_{\text{surf}}$ between surface layer and vacuum, and the depth of the impurity $d$ which only takes discrete integer values (Fig.~\ref{fig-app:SketchExpPW}). 

\begin{figure}
	\includegraphics[width=8.6cm]{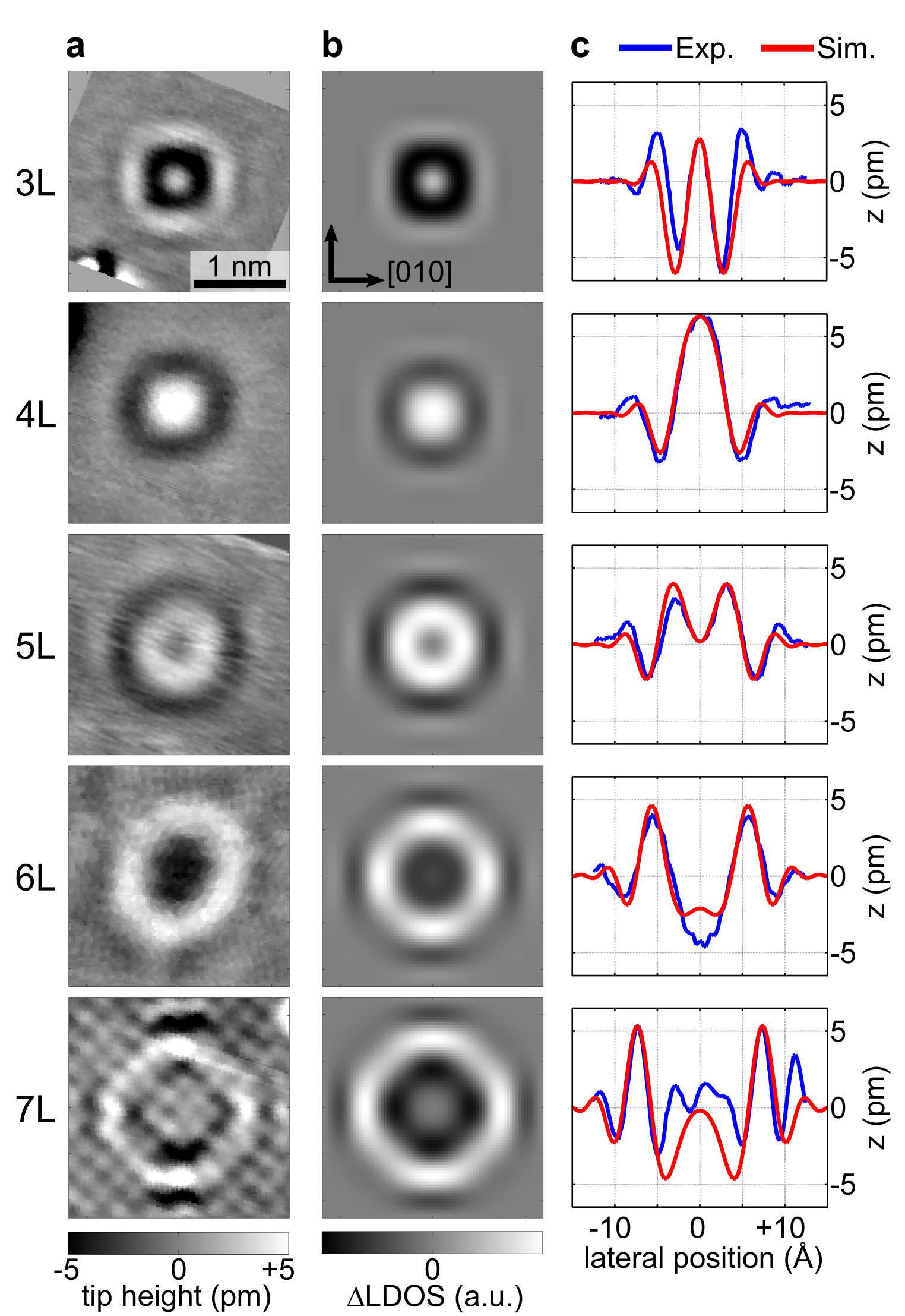}
	\caption{\label{fig-app:Ag-TB} Ag impurities in Cu(100) in various depths. Experiment and simulations show excellent agreement. (a) Experimental topographies ($2.4\times2.4$\,nm$^2$) at set points ($U=-10$\,mV / $I$(3L,7L)=2.0\,nA, $I$(4L,5L)=3.0\,nA, $I$(6L)=0.8\,nA) show for every layer their own characteristic shape increasing in size. For the 7th ML the focusing signal is superimposed with atomic resolution. Also, the center above the impurity is not perturbed as the focusing angle mainly aims in a direction of about 30$^\circ$ with respect to the surface normal. (b) TB-simulated topographies for fitted values of $\eta$ (varying between $\eta$=1.24$\pi$... 1.42$\pi$) for set-point dependent height ($h=5$\,\AA). (c) Comparison of experiment (blue) and simulation (red) for horizontal section ([010] direction) along the impurity.}
\end{figure}

\begin{figure}
	\includegraphics[width=8.6cm]{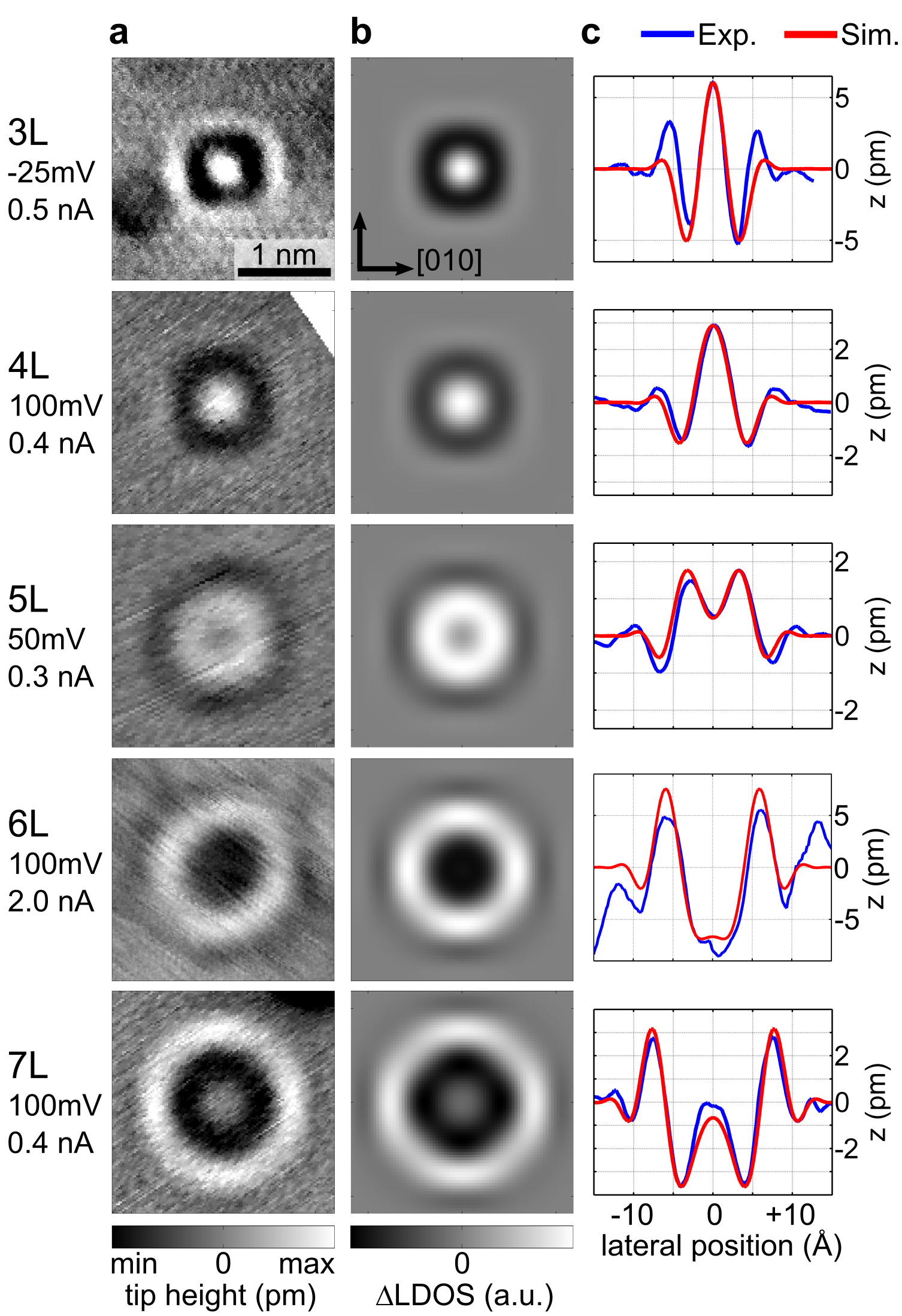}
	\caption{\label{fig-app:Ge-TB} Ge impurities in Cu(100) in various depths. Experiment and simulations show excellent agreement. (a) Experimental topographies ($2.4\times2.4$\,nm$^2$) at different set points. (b) TB-simulated topographies for fitted phase shift values ($\eta$=\{1.19,1.46,1.27,1.07,1.11\}$\pi$) for set-point dependent height ($h=7$\,\AA). (c) Comparison of experiment (blue) and simulation (red) for horizontal section ([010] direction) along the impurity shows excellent agreement.}
\end{figure}

The typical distances in scanning tunneling microscopy are about 5\,\r{A} to 8\,\r{A} where shorter distances lead to more laterally high-frequency features because of larger transmission of $k_{||}$  \cite{Chen2007}. Taking the bias voltage and current set points into consideration, we choose $h=5$\,\r{A} for Ag and $h=7$\,\r{A} for Ge data sets which proves to reproduce well the general appearance of interference patterns (cf. Figs.~\ref{fig-app:Ag-TB} and \ref{fig-app:Ge-TB}). The tip-sample distance $h$ does not have strong influence on the simulated topographies as a large change of $h$ only leads to a change of roughly $0.1\pi$ in phase shift which is about the size of the error bar for the extracted phase shift.

The surface layer spacing is set to $a_{\text{surf}}=a_{[100]}=181$\,pm in a rough approximation of the crystal potential  \cite{Chulkov1999}. Small changes in depth can be compensated by a shift in effective phase shift $\eta_{\text{imp}}$ because the phase contribution of additional propagation distance (Fermi wave length $\lambda_{\text{Cu}}=4.6$\,\AA) cancels with the new $\eta_{\text{imp}}$. A shift of 0.1\,ML in depth corresponds to approximately $-0.14\pi$ in phase shift. Therefore, as the parameters of the TB model are not independent of each other, there is not exactly one resulting parameter set. However, the lateral size of the surface signature which has to match with the focusing directions confines this flexibility to very few monolayers. 

The depth of an impurity $d$ is deduced from images with atomic resolution (for more information see Supplementary Note S2). The depth is restricted to discrete integer values as impurities only sit substitutionally in Cu atomic lattice points.

The effective phase shift in the TB model includes all phase contributions within the solid, even if they do not occur at the impurity. Firstly, this is a phase shift from the transmission from the vacuum to the metal crystal \cite{Woodruff1985}. Secondly, unlike the phase shift in the KKR framework, the TB model assumes propagation within the impurity cell which leads to a phase contribution. These contributions are identical for both impurity species, so the phase shift difference between them is independent of the contributions' absolute values. 
In Fig.~\ref{fig-app:Ag-TB} the surface signatures of Ag impurities from 3rd to 7th ML in Cu(100) are shown. The left column depicts the experimental topographies already presented in Fig.~\ref{fig:AgGeExpComp}b in the main text. With increasing depth, the picometer-high corrugations increase in lateral size. Each four-fold pattern has its characteristic shape of maxima and minima due to constructive and destructive interference. The center column (Fig.~\ref{fig-app:Ag-TB}b) depicts simulated topographies calculated with the TB model with a best-fit scattering phase, reproducing their experimental counterparts in size and shape. The sections (Fig.~\ref{fig-app:Ag-TB}c) along the [010] direction also indicate accurate agreement between experimental and simulated data. The deviations for the 7th ML originate from the atomic resolution superposed with the bulk focusing signal. We find an average effective scattering phase shift of $\eta_{\text{Ag}} =(1.32 \pm 0.04) \pi$.

Figure~\ref{fig-app:Ge-TB} shows the surface signatures of another non-magnetic impurity, Ge. Figure~\ref{fig-app:Ge-TB}a shows experimental topographies in monolayers 3-7. The simulated 2D TB topographies (Fig.~\ref{fig-app:Ge-TB}b) and sections of experiment and calculation (Fig.~\ref{fig-app:Ge-TB}c) show good agreement. We obtain an average effective scattering phase shift of $\eta_{\text{Ge}} =(1.23 \pm 0.05) \pi$.
The tight-binding model proves to be a reliable tool to describe the surface patterns of subsurface, non-magnetic impurities. This can be especially useful when ab-initio calculations are not available or for deeper layers where computational cost increases strongly for DFT.

\section{Electron-phonon coupling toy model \label{app:EPH1Dmodel}}
In the implementation of electron-phonon coupling in the tight-binding model, as described in the main text, we have only included the real part of the self-energy, which is responsible for the shift in center energy in the spectral function with respect to the single-particle system. Here, we analyze the role of the imaginary part for the spectral function in real space using a simple 1D toy model.

We approximate the quadratic dispersion of a free electron to be linear in the considered energy interval around the Fermi energy ($E_F = $7\,eV). The interaction with Debye phonons is introduced via a self-energy calculated from an Eliashberg function $\alpha^2F$ similar as described in the main text.

\begin{figure*}
	\includegraphics[width=17.2cm]{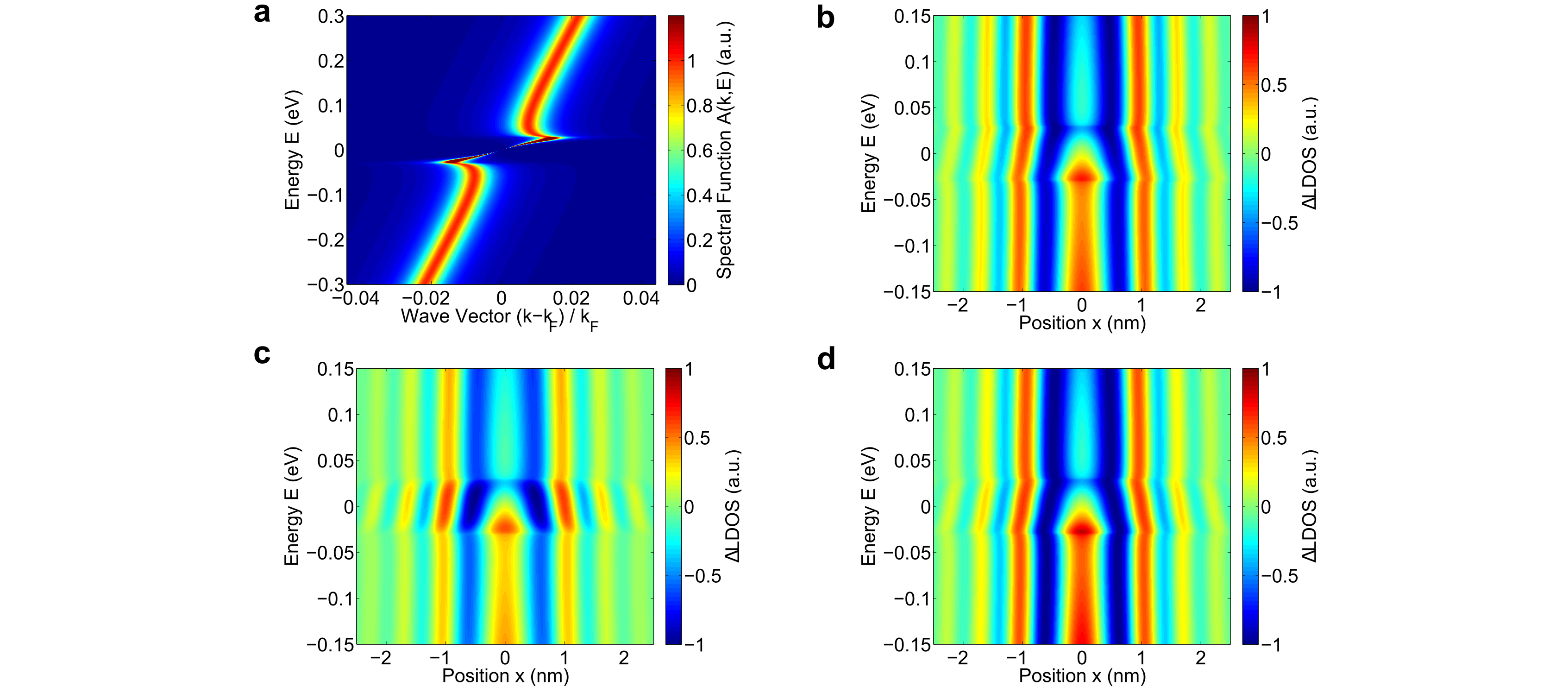}
	\caption{\label{fig-app:EPH1Dmodel} Role of the imaginary part of the self-energy in a one-dimensional toy model ($\lambda=5$, $\omega_D=$30\,meV). (a) Spectral function in momentum space shows characteristic kink. (b,c) Real space STS pattern for impurity ($d=$7ML, $\eta=1.0\pi$) with (b) only real part (c) both real and imaginary part of the self-energy. (d) Same as (c) with pattern normalized for each energy.}
\end{figure*}

The resulting renormalized spectral function in momentum space is shown in Fig.~\ref{fig-app:EPH1Dmodel}a for $\lambda=5.0$. The typical kink at the Debye energy $\hbar\omega_D=$30\,meV as well as the broadening due to quasiparticle interactions for energies $E>0$\,eV are found as known from the literature \cite{Grimvall1981}.

Via Fourier transform, we obtain the real space representation of the Green's function propagator and calculate a simulated STS pattern for an impurity in 7ML with effective scattering phase $\eta=1.0\pi$. Because the Friedel oscillations' amplitude in the 1D model is constant even for large distances, we include an additional damping term $\sim r^{-3}$ (with distance $r$ from impurity to surface position) into the LDOS. This accounts for a 3D propagation from impurity to the surface like in the experimental patterns.

For a propagator with only the real part of the self-energy, as implemented in the tight-binding model, we obtain a pattern of constant intensity with energy with a characteristic kink around the Fermi level (Fig.~\ref{fig-app:EPH1Dmodel}b). As this toy model is spatially isotropic and, thus, does not favor a certain electron focusing angle, there are multiple oscillations in the surface pattern. The smaller amplitude with larger distance only results from the additional 3D damping term.

Including the full self-energy (Fig.~\ref{fig-app:EPH1Dmodel}c), the imaginary part reduces the signal strength due to the finite lifetime. For $E>\omega_D$, the amplitude is attenuated by approximately 50\% while close to the Fermi level, the patterns remains almost unchanged. The whole pattern shape (i.e.\ the position of minima and maxima) is almost identical to the case with only $\Re e(\Sigma)$, as depicted in Fig.~\ref{fig-app:EPH1Dmodel}d where for each line the spectral function is normalized to 1.

From the study of the toy model, we conclude that our implementation of many-body effects into the Cu tight-binding model can describe the basic effects of electron-phonon coupling for the STS spectra. We do not claim exact quantitative accuracy for the tight-binding model. After all, despite describing well the overall topographic and spectroscopic signatures for different impurity species, it is a simple model that does not include all effects of the solid.

\begin{acknowledgments}
We gratefully acknowledge fruitful discussions with F.\ Heidrich-Meisner. Moreover, we acknowledge funding by the Deutsche Forschungsgemeinschaft through Grant No.\ LO1659/5-1 and WE1889/8-1 and funding from the European Research Council (ERC) under the European Union's Horizon 2020 research and innovation programme (ERC consolidator grant 681405 'DYNASORE').
\end{acknowledgments}

\end{document}


\title{Supplementary Material: Scanning Tunneling Spectroscopy of Subsurface Ag and Ge Impurities in Copper}

\author{Thomas Kotzott}
\affiliation{IV. Physikalisches Institut -- Solids and Nanostructures, Georg-August-Universit\"at G\"ottingen, 37077 G\"ottingen, Germany}

\author{Mohammed Bouhassoune}
\affiliation{Peter Gr\"unberg Institut and Institute for Advanced Simulation, Forschungszentrum J\"ulich and JARA, 52425 J\"ulich, Germany}
\affiliation{D\'epartement de Physique, FPS, Cadi Ayyad University, Marrakech, Morocco}

\author{Henning Pr\"user}
\affiliation{IV. Physikalisches Institut -- Solids and Nanostructures, Georg-August-Universit\"at G\"ottingen, 37077 G\"ottingen, Germany}

\author{Alexander Weismann}
\affiliation{IV. Physikalisches Institut -- Solids and Nanostructures, Georg-August-Universit\"at G\"ottingen, 37077 G\"ottingen, Germany}
\affiliation{Institut f\"ur Experimentelle und Angewandte Physik, Christian-Albrechts-Universit\"at zu Kiel, 24098 Kiel, Germany}

\author{Samir Lounis}
\affiliation{Peter Gr\"unberg Institut and Institute for Advanced Simulation, Forschungszentrum J\"ulich and JARA, 52425 J\"ulich, Germany}
\affiliation{Faculty of Physics, University of Duisburg-Essen and CENIDE, 47053 Duisburg, Germany}

\author{Martin Wenderoth}
\email[]{martin.wenderoth@uni-goettingen.de}
\affiliation{IV. Physikalisches Institut -- Solids and Nanostructures, Georg-August-Universit\"at G\"ottingen, 37077 G\"ottingen, Germany}

\date{October 12, 2021}

\maketitle

\section{Surfaces of dilute Cu alloys \label{app:surfacedilutealloys}}
The samples, Cu(100) with dilute Cu alloys in the topmost layers, were prepared in a home-built UHV preparation chamber with a base pressure of $p = 5\times10^{-11}$\,mbar. The Cu(100) single crystals were cleaned by cycles of sputtering by Ar$^+$ ions with an energy of 700\,eV and annealing to 680\,K. In the last cycles the sputtering and annealing times were reduced and the temperature was lowered to 630\,K in order to flatten the surface without causing segregation of bulk defects towards the surface. The surface quality was checked by low-energy electron diffraction, Auger electron spectroscopy and STM which presents the sample as clean with terraces of a width of up to a few 100\,nm. Dilute alloys in the 20 topmost layers are fabricated by co-deposition of 20\,ML copper and $<1\%$ silver and germanium, respectively, by two electron beam evaporators. Cu is evaporated at 1\,ML/min. In order to achieve very small ratios of composition, the impurity evaporator is equipped with a stepper motor which allows shutter opening times of 100\,ms. During evaporation, the sample is held at a temperature of T$\approx$100\,K. In order to restore a flat surface that can be investigated by STM, the sample is flashed to $T=520$\,K for 3 seconds after growth. Subsequently, it is immediately transferred into the low temperature STM.

\begin{figure}
	\includegraphics[width=10cm]{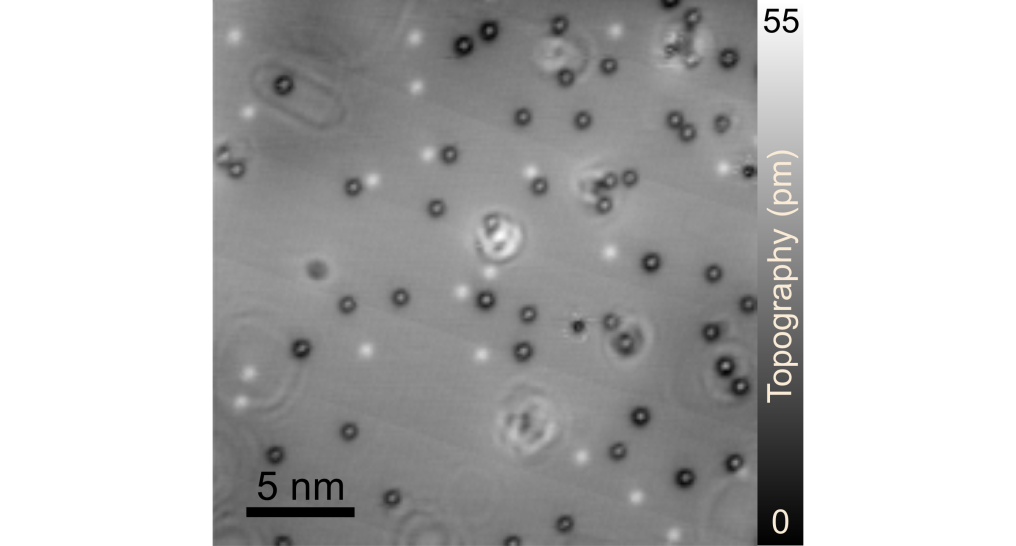}
	\caption{\label{fig-app:LargeTopoAg}Topography ($25\times25$\,nm$^2$, $U = 100$\,mV / $I = 2$\,nA) with various patterns in the LDOS at the surface due to Ag impurities. Bright spots with dark rings correspond to second layer impurity atoms. On the left, the signatures of more deeply buried, single impurities are found.}
\end{figure}

The surfaces of dilute Cu alloys show different patterns at the surface corresponding to different defects. An example of a topography with Ag impurities is shown in Fig.~\ref{fig-app:LargeTopoAg} which is similar to the large-scale topography for Ge presented in the main text (cf. Fig.~1a). Many second layer Ag impurities are visible as strong contrast white dots with surrounding black rings. In the left center two impurities are buried, one rather close to the surface and one deeper impurity with clear four-fold pattern. Additionally, signatures of clusters presenting richer patterns are found (e.g.\ center, bright white contrast) which can either be linked to subsurface impurity clusters or nanocavities of residual argon gas from the sputtering process.

\begin{figure*}
	\includegraphics[width=17.2cm]{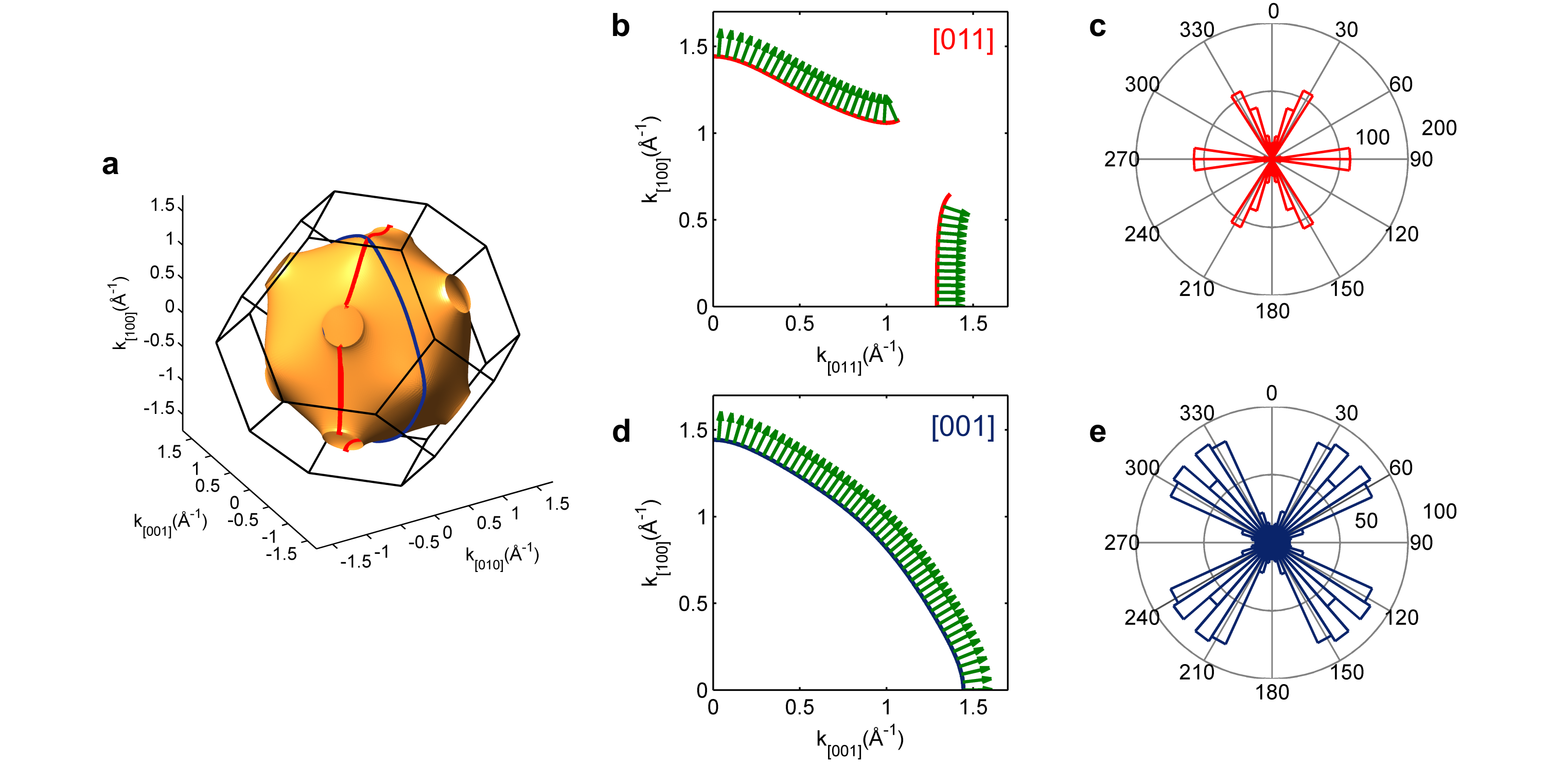}
	\caption{\label{fig-app:ElectronFocusing}Electron focusing directions for Cu(100). The group velocity of electrons is given by the gradient of the band structure. For flat regions on the Fermi surface, many electrons propagate in the same direction. (a) Three-dimensional representation of the Cu Fermi surface with indications in red and blue for the [011] and [001] directions, respectively. (b,d) Sections of the Cu Fermi surface for the indicated directions, respectively. Green arrows depict the group velocities. (c,e) Polar histograms of focusing directions for the Fermi surface sections in (b) and (d), respectively. For the [001] direction, the dominating focusing angles are around 30$^\circ$ and 60$^\circ$.}
\end{figure*}

The four-fold signature for single, buried impurities are the consequence of electron focusing. Electrons within the crystal propagate corresponding to their group velocity which is given by the gradient of band structure \cite{Weismann2009}. Hence, for flat regions on the isoenergy contour (e.g.\ the Fermi surface for $E=0$\,eV), there is strong focusing in the corresponding direction, while for regions with high curvature the group velocity vectors point in different directions, leading to a weak signal at the surface. This effect is depicted for two sections of the Cu Fermi surface in Fig.~\ref{fig-app:ElectronFocusing}. The electrons dominantly propagate in directions where group velocity vectors accumulate, leading to surface interference patterns above buried impurities. In semiconductors or other materials with comparably low charge carrier density, subsurface dopant atoms can be mapped by STM in a different mechanism. The donor's extended Coulomb potential around the impurity site can be modified by switching the dopant's charge state by tip-induced band bending \cite{Teichmann2008,Song2012}. This mechanism is excluded for the impurities investigated here, buried in the metal Cu. The charge carrier density is significantly higher and the Thomas-Fermi screening length as short as 0.55\,\AA, so that any charged region is screened on short length scales. 

\begin{figure}
	\includegraphics[width=10cm]{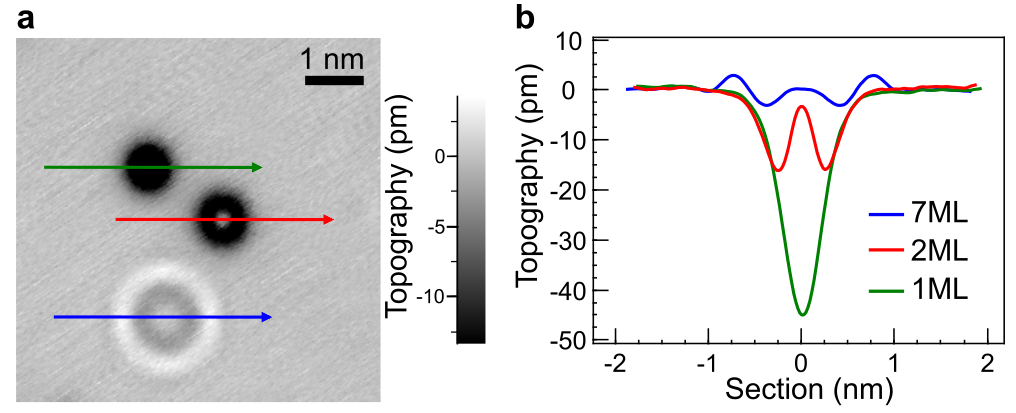}
	\caption{\label{fig-app:SurfaceLayers}Impurities at the surface ($6.1\times6.1$\,nm$^2$, $U=100$\,mV, $I=0.4$\,nA). Surface layer Ge impurity ('1ML', green) shows a deep protrusion and the second ML impurity (red) one oscillation as visible in Fig.~\ref{fig-app:LargeTopoAg}. For comparison, a bulk 7th ML impurity (blue) shows low picometer-high contrast with larger lateral extent.}
\end{figure}

In Fig.~\ref{fig-app:SurfaceLayers} three different types of single impurities in Cu(100) are shown in topography and line sections. The signature of first monolayer impurities is a depression for Ge (and also previously, e.g.\ Co \cite{Kloth2014}) with an amplitude of typically 30\,pm. Second ML defects, only covered by one monolayer of Cu, reveal a bright spot surrounded by a dark ring. The electronic contrast is typically 10--15\,pm. As a reference also a 7th ML impurity with picometer-high signature is included in the data set. Due to very good imaging conditions in the presented data set, the typical heights for the features are exceeded. 

\section{Identification of single subsurface impurities \label{app:singlesubsurfaceimps}}
\subsection{Determination of impurity depth}
\begin{figure}
	\includegraphics[width=8.6cm]{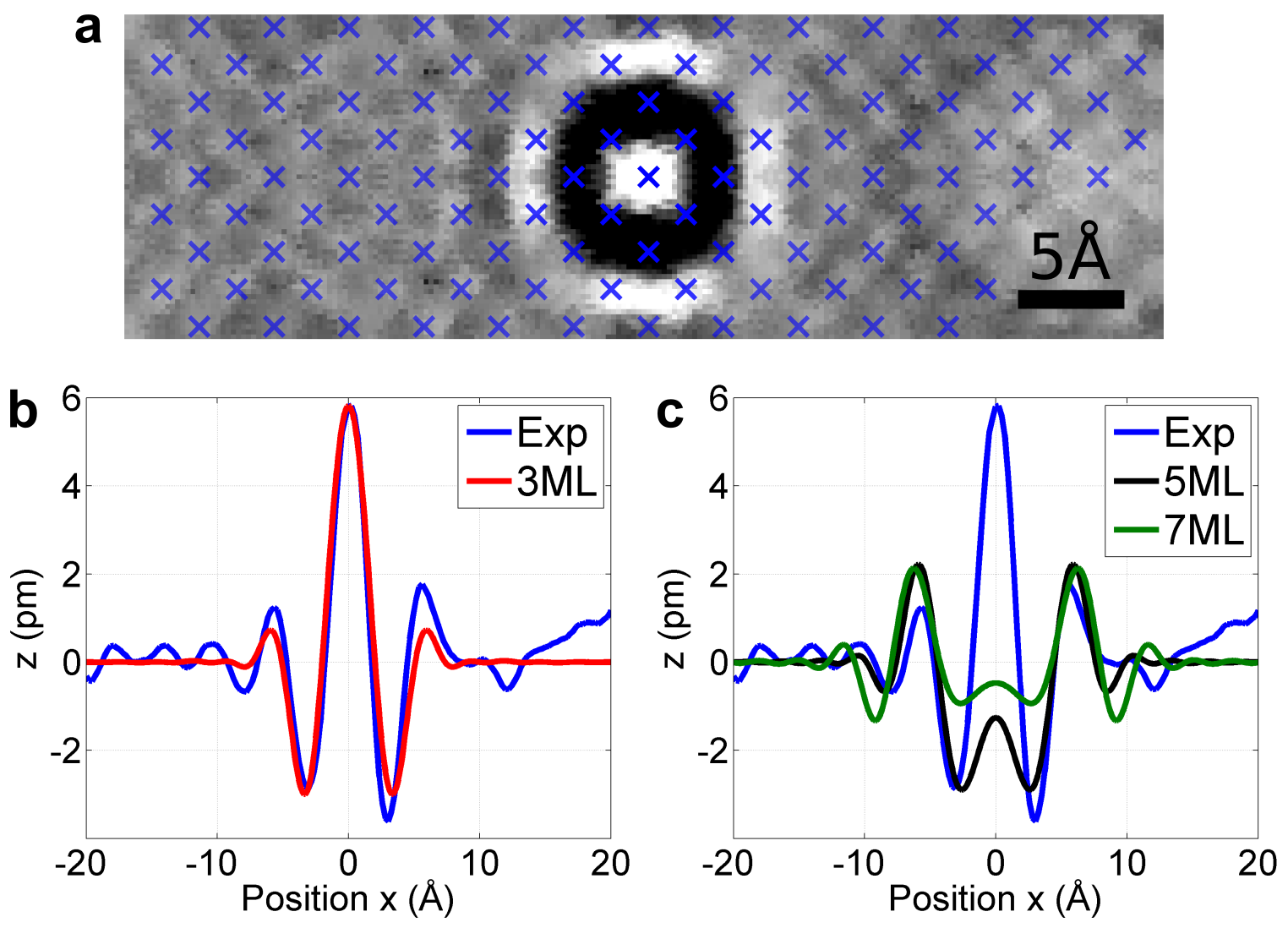}
	\caption{\label{fig-app:AtomicResDepth}Determination of impurity's depth by atomic resolution data. (a) Surface signature of Ge impurity buried in the 3rd ML in Cu(100) superimposed with atomic resolution. The surface lattice is labelled with blue crosses and continued across the interference pattern. The atomic resolution allows to deduce the depth of the impurity. The pattern is always centered around the impurity atom. Here, it is also centered around a surface atom. Thus, the buried impurity sits perpendicular to the surface in line under this surface atom. Therefore, due to the fcc structure of Cu, the impurity has to be located in an odd ML (with the surface layer labelled as first layer). (b) Section along [010] direction across the interference pattern for experiment (blue) and tight-binding simulation (3ML, $\eta$=1.25$\pi$). (c) Comparison of experimental section and best-fit tight-binding simulations for 5ML ($\eta=0.6\pi$) and 7ML ($\eta=1.7\pi$). The pattern's lateral size is too small for a 5th ML or even 7th ML impurity, hence, this impurity is located in the 3rd ML.}
\end{figure}

The depth of an impurity is deduced from STM topographies with atomic resolution. The center position of the surface interference pattern with respect to the surface lattice determines if the impurity is located in an odd or even crystal monolayer. The surface pattern is always centered around the buried atom due to symmetry. If the surface signature coincides with a surface atom, then the impurity sits perpendicular to the surface below this surface atom. Hence, with Cu being an fcc single crystal, the impurity is located in an odd monolayer. Here, the surface layer is counted as first layer. If the surface signature's center is located between surface atoms, then the impurity is positioned in an even crystal monolayer. An example for such an analysis is shown in Fig.~\ref{fig-app:AtomicResDepth}a. The surface interference pattern of this impurity is centered around a surface atom (marked with a cross), so the impurity is located in an odd crystal monolayer.

Additional information for the determination of an impurity atom's depth is taken from the lateral pattern size which is restricted by the focusing angle, i.e., the main directions of electron focusing. Because of this, for a full data set in which all depths are available, the patterns can be ordered by size and then assigned a crystal monolayer. Alternatively, tight-binding calculations (see Appendix A in the main text) can be performed for different depths, for each fitting the best effective scattering phase shift. As the impurity positions are limited to a discrete number of crystal layers and by atomic resolution data further reduced to only even or odd layers, the comparison of simulation and experimental data reveals the impurity's depth. For the exemplary depth analysis in Fig.~\ref{fig-app:AtomicResDepth}, in (b) the experimental data is compared with a tight-binding simulation for a third monolayer ('3ML') impurity. The patterns match far better than the candidates of 5th ML and 7th ML in subfigure (c). For deeper layers, the main weight of the interference pattern is shifted towards larger distances, whereas for the 3rd ML the main signal is located in the center peak.

For data sets where no atomic resolution was obtained, the depth is determined by comparison with tight-binding simulations of various depth with the average effective scattering phase shift, assuming a constant scattering phase shift for all bulk layers. Once the layer is identified (e.g.\ comparison of lateral size, ratios in contrast pattern), in a next step an effective scattering phase shift is fitted more precisely with knowledge of the impurity's depth.

\subsection{Single impurities in bulk Cu(100)}
The scattering centers are identified as point-like, single atoms of only one species. The key for this is the control of the growth process which ensures that the samples are prepared cleanly with only the impurities' species as contamination.
Reference measurements with the clean Cu(100) sample and a Cu(100) sample with additional evaporated Cu layers are inconspicuous. No argon cavities as result of preparation are found, as reported by Schmid et al, Kurnosikov et al.\ and others \cite{Kurnosikov2008,Kurnosikov2009,Schmid2000}. The interference patterns for residual sputter gas (Ar, Ne) can be distinguished from the patterns used in our analysis as they show richer surface interference patterns with larger lateral extent (e.g.\ in Fig.~\ref{fig-app:LargeTopoAg}).

For the dilute alloys, the concentration of impurities is usually $<1\%$, so that impurity atoms can hardly form clusters. In some topographies we find clusters of Ag or Ge with rich surface signatures which are easily distinguished from point-like scattering centers which show clearly defined four-fold interference patterns confined to approximately 1.5 Friedel oscillations \cite{Weismann2009}. 

Nevertheless, the tip quality is crucial for resolving the surface signatures of buried impurities. They show apparent heights of few picometers in topography which correspond to LDOS modulations of around only 1\% in spectra. Data sets with reproducible, well-defined interference patterns as shown in the main text can be recorded with a suitable sharp tip. For dull tips, we obtain broader features which make it more difficult to distinguish between subsurface structures and other defect contrasts, which is why we exclude these from the analysis.

\clearpage
\section{Ab-initio calculations and scattering phase shifts \label{app:ab-initio}}
The ab-initio calculations are performed using DFT as implemented in the full potential Korringa-Kohn-Rostoker Green function (FP-KKR-GF) method \cite{Papanikolaou2002} with the local density approximation LDA \cite{Vosko1980}. First, the electronic structure of 18 layers fcc Cu slab with two additional vacuum regions (3 vacuum layers on each side of the Cu slab) are taken along the (001) direction. The decimation technique \cite{Garcia-Moliner1986,Szunyogh1994} is adopted to simulate the semi-infinite substrate to avoid size artifacts in the charge density. The experimental lattice parameter $a = 3.61$\,\r{A}  was considered without surface relaxations, which are negligible. Then each impurity is embedded underneath the surface together with its neighboring atoms, defining a cluster of atoms, where the charge is allowed to be updated during self-consistency. The induced charge density is then computed in the vacuum at $h = 3.61\,\text{\AA}$ above the surface, proportional within the Tersoff-Hamman approach \cite{Tersoff1983} to the tunneling signal measured with STS. While the self-consistent calculations required 40 $\times$ 40 $k$-points in the two-dimensional Brillouin zone, the theoretical STS spectra were obtained with a set of 200 $\times$ 200 $k$-points.

The calculations are performed to obtain simulations of the experimental spectroscopy data and energy-dependent scattering phase shifts at the different impurity species. As the former results are discussed in the main text, here we mainly focus on the phase shifts at the Fermi energy.

We obtain scattering phases $\delta$ for host and impurity species consistent with previous studies \cite{Braspenning1982}. The values of phase shifts at the Fermi energy are listed in Tab.~\ref{tab:DFT-phaseshifts}. A comparison between 5th ML and 3rd ML shows only tiny changes for the different depth. This indicates a depth-independent scattering phase shift which supports the assumption in our analysis for extracting the phase shift from experimental data.

\begin{table}
	\caption{\label{tab:DFT-phaseshifts}
		Scattering phase shifts $\delta$ from ab-initio calculations at Fermi energy $E_F$ for different species in 3rd and 5th ML. 
	}
	\begin{ruledtabular}
		\begin{tabular}{lccc}
			\textrm{Species/Layer}&
			\multicolumn{1}{c}{\textrm{$s$}}&
			\multicolumn{1}{c}{\textrm{$p$}}&
			\multicolumn{1}{c}{\textrm{$d$}}\\
			\colrule
			Cu 3ML & -0.086 & 0.070 & -0.180\\
			Cu 5ML & -0.084 & 0.074 & -0.180\\\hline
			Ag 3ML & -0.20 & -0.04 & -0.22\\
			Ag 5ML & -0.20 & -0.01 & -0.22\\\hline
			Ge 3ML & 0.87 & 0.85 & 0.045\\
			Ge 5ML & 0.87 & 0.86 & 0.044\\
		\end{tabular}
	\end{ruledtabular}
\end{table}

With the tight-binding model (c.f. Appendix A in the main text), we parametrize the experimental surface patterns with a single effective scattering phase shift $\eta_{\text{imp}}$ which describes the phase relation of incoming and scattered wave. This parameter is determined by the scattering potential of the impurity. The corresponding quantity used in quantum scattering theory is the scattering phase shift $\delta$ which is linked to the scattering matrix $t$ by $t=-(1/\sqrt{E}) \sin(\delta) \exp(i\delta)$. In a Cu crystal with Bloch electron states, the scattering amplitude and phase due to an impurity are given by the complex number $\Delta t= t_{\text{imp}} - t_{\text{Cu}}$ which is the difference of the impurity scattering matrix with respect to the host  \cite{Lounis2011}. 

The experimentally determined effective scattering phase shift $\eta_{\text{imp}}$ corresponds to arg($\Delta t$). The scattering amplitude, i.e., the amplitude of the complex scattering matrix difference, is contained in the height of electronic contrast measured by STM. However, electronic contrast is very difficult to quantify as transmission strongly depends on absolute tip-sample distance and tip shape. In the calculations, Ge shows as an impurity in Cu a three to four times higher scattering amplitude for $s$ and $p$ orbitals than the $d$ orbital contribution. For Ag, the orbital components only vary by a factor of two, with all amplitudes being half of the Ge's value for $d$ orbitals or less.

\begin{figure}
	\includegraphics[width=15.0cm]{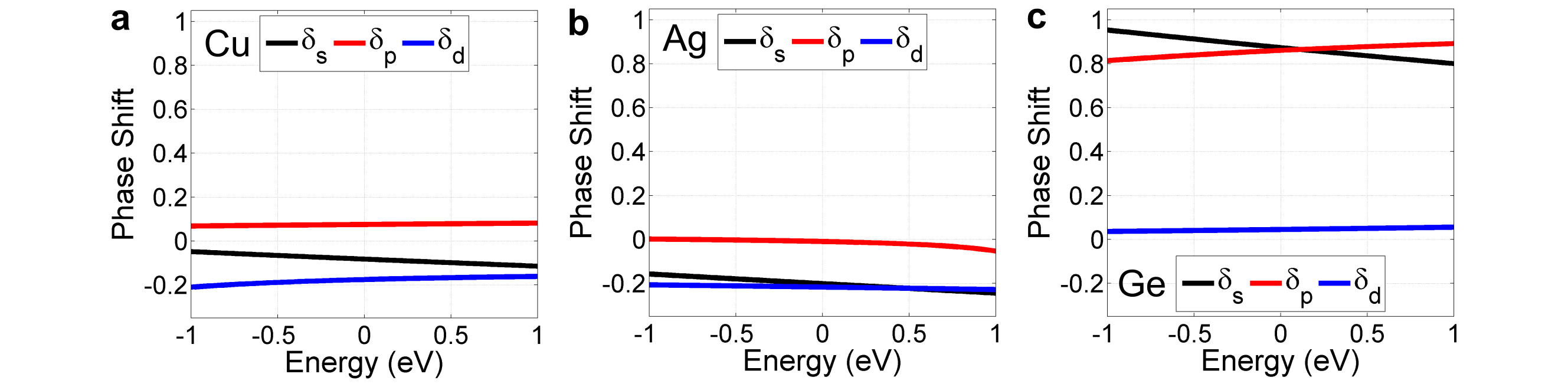}
	\caption{\label{fig-app:DFTPhaseShifts}Energy-dependent phase shifts determined by ab-initio calculations. The phase shift is plotted for the 5th ML for orbital contributions $s$ (black), $p$ (red) and $d$ (blue) for species (a) Cu, (b) Ag and (c) Ge. The phase is approximately constant with respect to energy around $E=0$\,eV which supports the approximation of an energy-independent phase shift in the analysis with the tight-binding model. Scattering properties of impurities in Cu are given by the difference of scattering matrices that are determined by the depicted KKR scattering phase shifts.}
\end{figure}

The energy-dependence of scattering phases for $s$, $p$, and $d$ orbital characters are shown in Fig.~\ref{fig-app:DFTPhaseShifts} for Cu, Ag, and Ge. The phase around the Fermi level is approximately constant. The strongest energy-dependence from $-1$\,eV to $+1$\,eV show the $s$ and $p$ orbitals of Ge. The scattering amplitude, i.e., the absolute value of difference of the complex scattering matrices of Ge and Cu, of these orbitals is 3-4 times higher than for the $d$ orbitals. 

\begin{figure}
	\includegraphics[width=8.6cm]{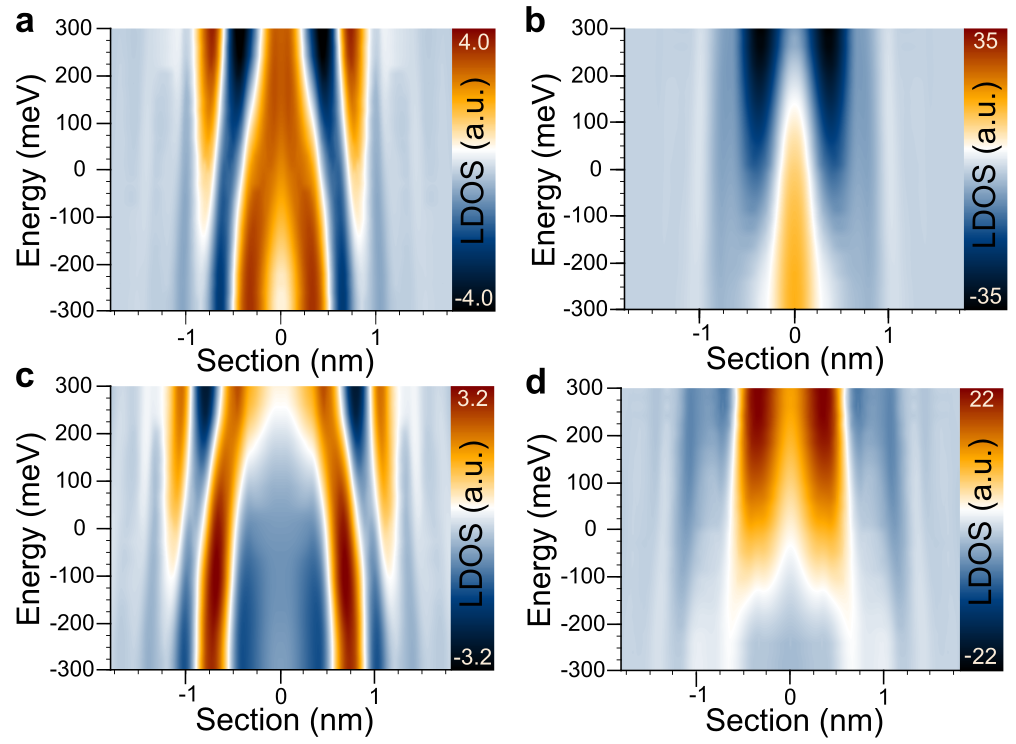}
	\caption{\label{fig-app:DFT-STSx4}Spectroscopic sections calculated by ab-initio KKR-DFT calculations compared for different depths and species. (a) 5th ML Ag impurity (b) 5th ML Ge impurity (as shown in Fig.~4b) (c) 7th ML Ag impurity (as shown in Fig.~4a) (d) 7th ML Ge impurity. Although the lateral size of interference patterns of one depth is similar for both species, the surface signatures differ in shape and LDOS amplitude. The LDOS amplitude is up to seven times higher for Ge. While Ag patterns show no strong change in scattering amplitude with energy, the Ge spectra are dominated by a contrast inversion in the center while only in the outer regions plain dispersive features (similar to Ag) are observed.}
\end{figure}

Resulting from the ab-initio calculations, we obtain simulated spectroscopic sections as shown in Fig.~\ref{fig-app:DFT-STSx4} which also comprises the data shown in the main text. While the lateral sizes of the surface signatures are similar for both species for a specific depth, Ag and Ge differ strongly in LDOS amplitude and pattern shape.

\begin{figure}
	\includegraphics[width=15cm]{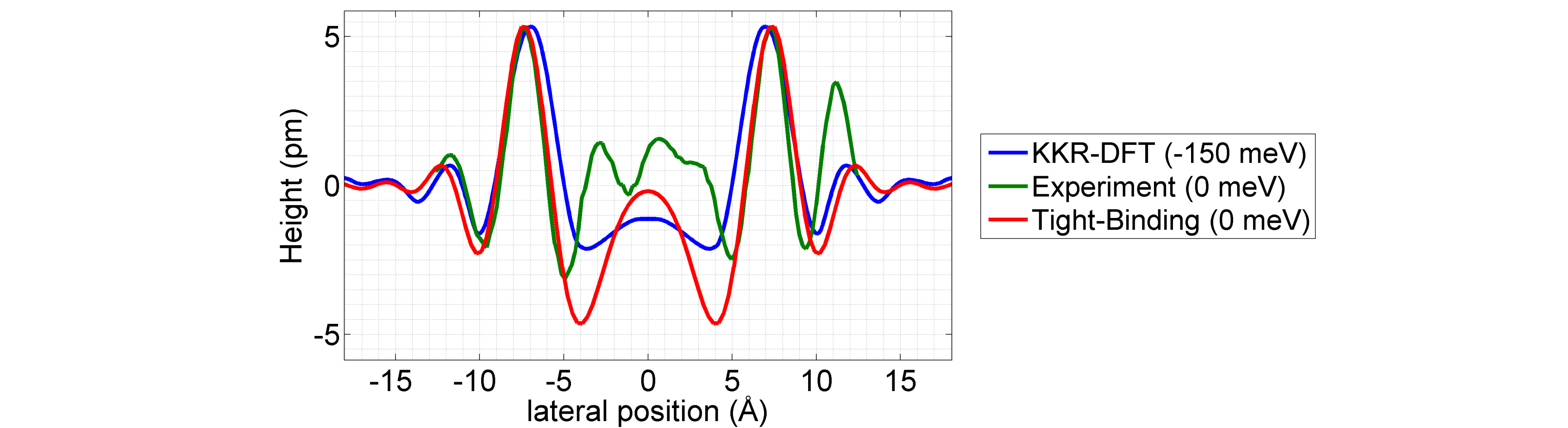}
	\caption{\label{fig-app:AgSectionsComparison} Comparison of ab-initio calculation (blue), experiment (green), and tight-binding model (red) for a Ag 7th ML impurity. For a quantitative match of the peak positions the ab-initio energy scale is shifted by $150$\,mV. The center part shows deviations in the experimental data due to the superposition of atomic resolution.}
\end{figure}

As discussed in the main text, for Ag we find good agreement between ab-initio calculations and the experimental data. In order to obtain a match for the overall shape and lateral size of the interference pattern, in the ab-initio calculations either the scattering phase shifts have to be slightly corrected or the energy scale has to be shifted by $150$\,mV. Such an energy shift is known from the literature to be possible for metallic electronic states as obtained from LDA with respect to those measured experimentally \cite{Rangel2012}. In Fig.~\ref{fig-app:AgSectionsComparison} the corresponding shifted sections are shown with good agreement of peak positions and ratio of peak intensities. The center part shows deviations due to the superposition of atomic resolution in the experimental data. For comparison, a section from a tight-binding simulation is also included.  Alternatively, if either the position of the impurity's electronic states or the atomic relaxation of Cu atoms around the impurity is underestimated, the scattering phase shifts are modified in the ab-initio calculations, possibly causing small differences in comparison with the experimental data.

\clearpage
\section{Deeply buried impurities \label{app:deeplyburied}}
\begin{figure*}[!b]
	\includegraphics[width=16.2cm]{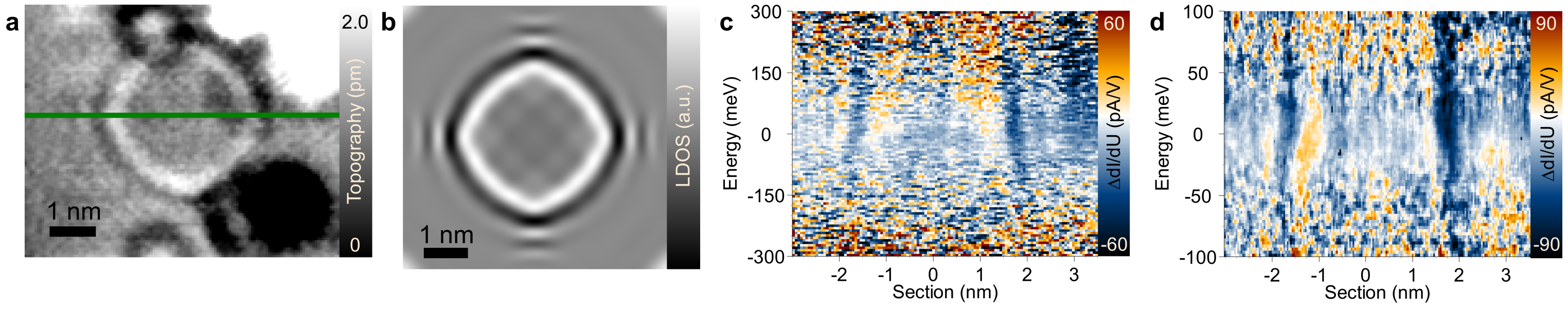}
	\caption{\label{fig-app:deeplyburied} Deeply buried Ge impurity in Cu(100) in the 18th monolayer. Topographic and spectroscopic contrast is detectable with good agreement to the TB-simulation despite large depth. (a) Topography ($7.0\times5.9$\,nm$^2$, $U=-100$\,mV / $I=0.5$\,nA) of an 18th ML Ge impurity. Adjacently, signatures of other buried impurities and surface defects are depicted, partially overlapping with it. Two intensity maxima and one minimum characterize the four-fold ring-like feature. (b) Tight-binding calculation of topography for the impurity ($6.0\times6.0$\,nm$^2$, $\eta=1.25\pi$). The simulation shows excellent agreement with the experiment. (c) Spectroscopic section along [010]-direction, labelled in green in (a). The difference $\Delta$d$I$/d$U$ with respect to clean Cu surface is plotted. The ring-like structures move towards the center for higher energies due to electronic dispersion. (d) Spectroscopic section similar to (c) with smaller energetic range. Around the Fermi level, an additional bending of the pattern is found that is assigned to electron-phonon coupling.}
\end{figure*}

Due to the electron focusing effect, the signature of buried impurities is detected by an STM despite the screening effects of the host's high charge density. While in previous data we have shown interference patterns corresponding to impurities of up to the 7th ML, in Fig.~\ref{fig-app:deeplyburied} we show a deeply buried Ge impurity. In Fig.~\ref{fig-app:deeplyburied}a, the topographic surface pattern of a 18th ML impurity is presented. The four-fold ring structure of more than 3\,nm diameter is clearly discriminable despite the adatoms and other buried impurities in the surrounding surface region. Because the focusing pattern is limited to distinct 1.5 wave lengths and clusters would lead to richer surface structures, we assign the pattern to a single impurity. We identify the impurity's depth by comparison with tight-binding simulations (Fig.~\ref{fig-app:deeplyburied}b) where we obtain good agreement.

In Fig.~\ref{fig-app:deeplyburied}c and Fig.~\ref{fig-app:deeplyburied}d, the spectroscopic signature is shown for two energy ranges. For $\pm$300\,mV (Fig.~\ref{fig-app:deeplyburied}c), the dominant feature is the shrinking of the ring-like structures for higher energies. This is understood by electronic dispersion. For high absolute values of energies, the signal is more difficult to detect as the tunnel current noise is also increasing. For these long propagation distances, there might be additional effects by electron-electron scattering for energies further from the Fermi level. As the impurity is located 30\,\AA\ below the crystal's surface, the electrons in the interference pattern have travelled coherently 7\,nm through the Cu crystal. In Fig.~\ref{fig-app:deeplyburied}d, we show a spectroscopic section for an energy range from $-100$\,mV to $+100$\,mV. As in the data before, we find an additional bending at the Fermi level due to electron-phonon coupling. Even for this depth, the renormalized band structure of interacting electrons is resolved in spectroscopy.

\clearpage

\section{Many-Body Effects \label{app:manybodysmall}}
\begin{figure*}[!h]
	\includegraphics[width=16.2cm]{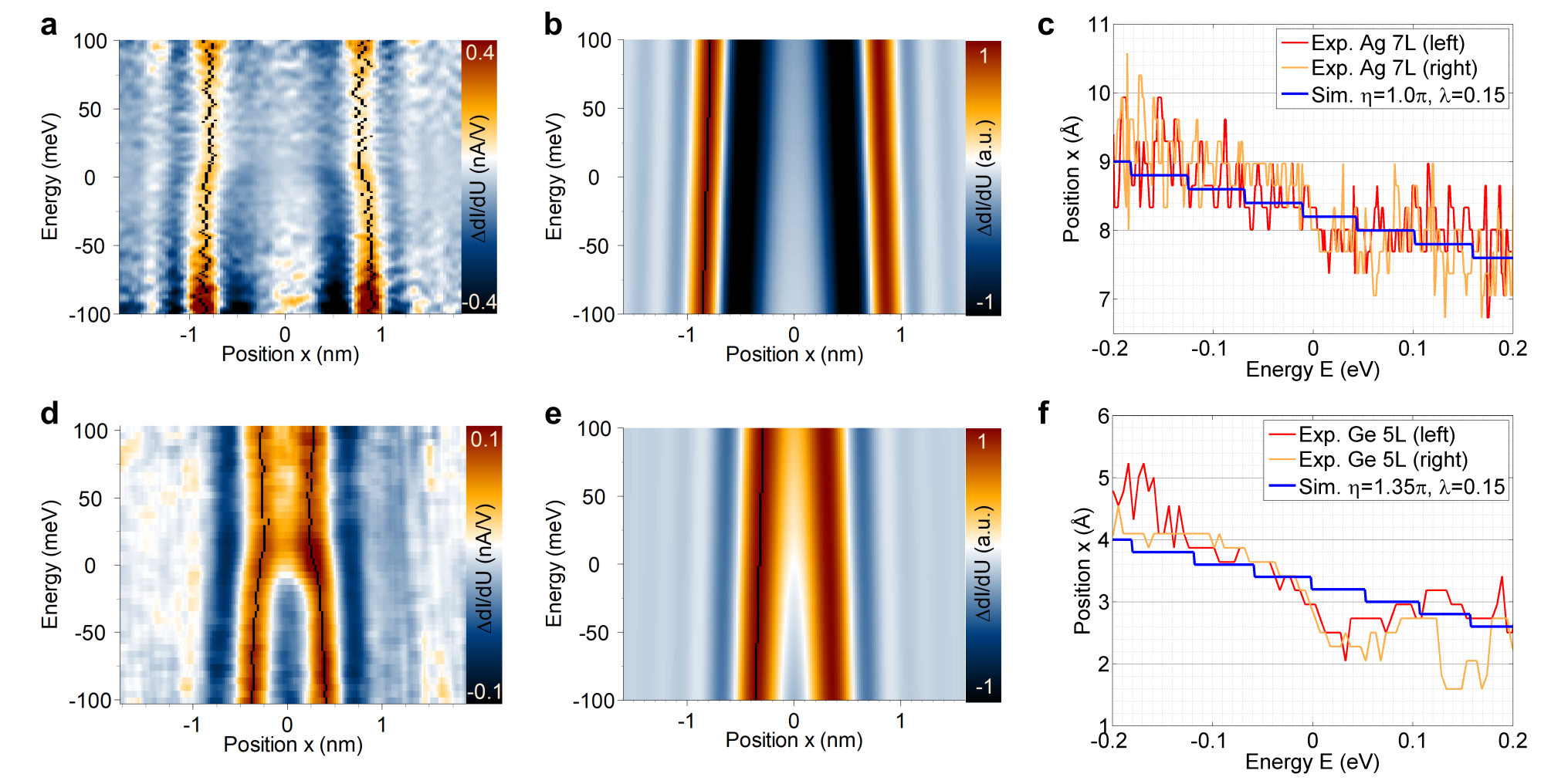}
	\caption{\label{fig-app:EPHAgGe0.15} Spectroscopic signatures of electron-phonon coupling with $\lambda=0.15$. (a,d) Experimental STS data for Ag 7ML ($U = -300$\,mV / $I = 2.0$\,nA) and Ge 5ML ($U = -300$\,mV / $I = 0.7$\,nA). The maximum of the pattern (left and right) is labelled (dark blue). (b,e) Tight-binding simulation for the respective impurities including electron-phonon coupling in the Debye model ($\lambda=0.15$, $\omega_D=30$\,meV, $h=5$\,\AA). (c,f) 1D Comparison of the maximum position in experimental (red/orange) and simulated (blue) data.}
\end{figure*}

In Fig.~\ref{fig-app:EPHAgGe0.15}, a compilation of experimental data and tight-binding simulations including many-body effects similar to Fig.~6 in the main text is shown, here with an electron-phonon coupling of $\lambda=0.15$. This corresponds to the bulk value of Cu \cite{Grimvall1981}. We find that the characteristics in the simulated spectra are so faint that they are not detected neither in the 2D maps nor the comparison plot. This shows that we need a value of $\lambda$ significantly higher than the known average bulk value in order to describe the features found in the experiment.

\clearpage

\section{Spectroscopic signatures and phonon spectrum \label{app:EPHthermalbroadening}}
\begin{figure*}[h]
	\includegraphics[width=16.0cm]{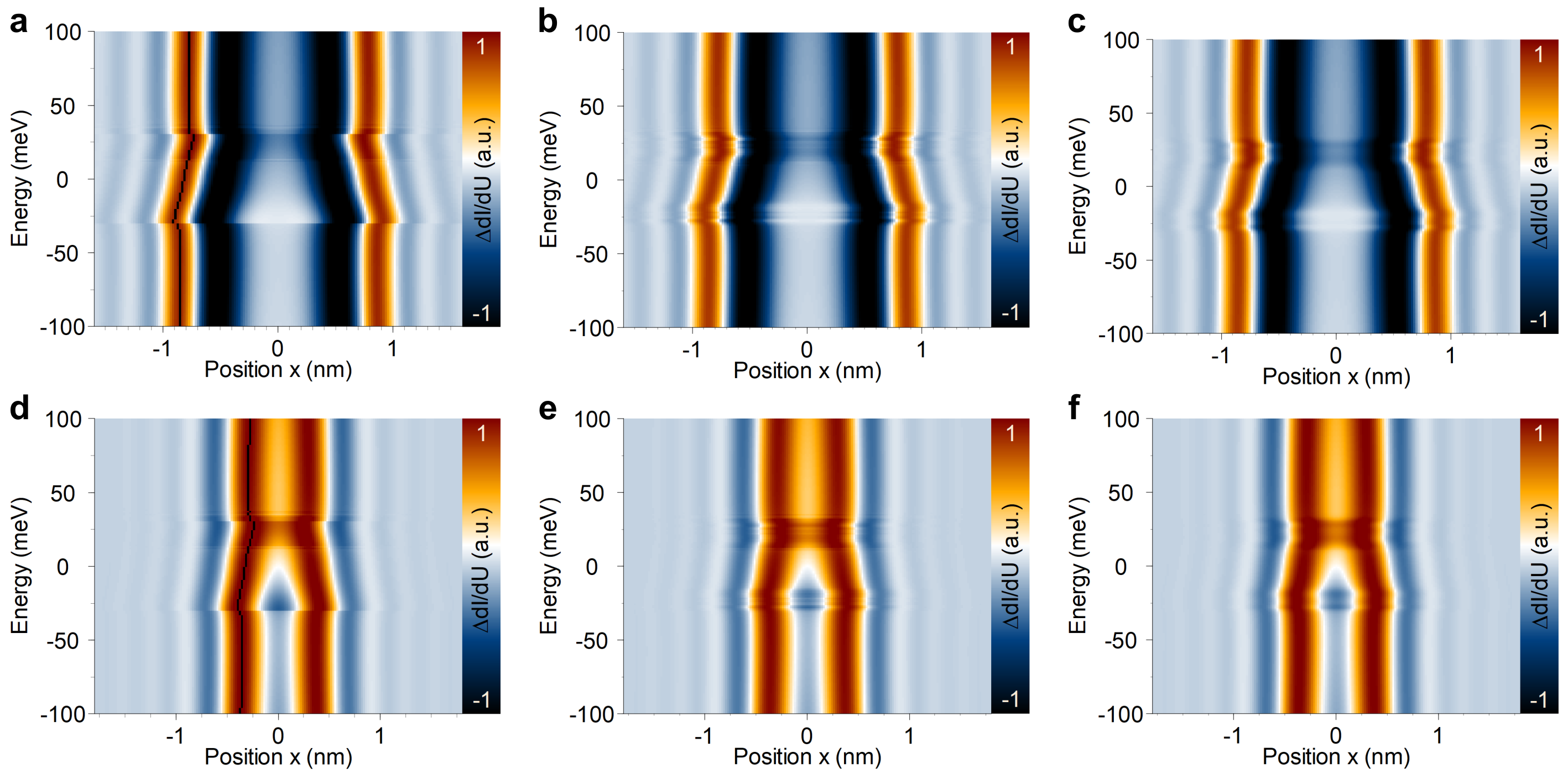}
	\caption{\label{fig-app:EPHThermalBroadening} Spectroscopic sections for (a-c) 7th ML Ag and (d-f) 5th ML Ge impurity, as calculated by tight-binding simulations including a self-energy ($\lambda=5.0$, $\omega_D=30$\,meV) comparing the Debye model with the Cu phonon spectrum without and with thermal broadening. (a,d) Self-energy based on Debye model, as shown in the manuscript in Figure 6. (b,e) Cu phonons, STS at $T$ = 0K. (c,f) Cu phonons, STS at $T$ = 6K.}
\end{figure*}

In Fig.~\ref{fig-app:EPHThermalBroadening}, a comparison of the spectroscopic signatures of a 7th ML Ag impurity and a 5th Ge impurity are shown for different phonon spectra. The tight-binding simulations including a self-energy based on the Debye model (Fig.~\ref{fig-app:EPHThermalBroadening}a,d) are also shown in Fig.~6 in the manuscript. Here, they are compared with calculations using a self-energy based on the Cu phonon spectrum. One finds that, firstly, the Debye model approximation of an abrupt cut-off at the Debye energy leads to discontinuous line shape. Therefore, the self-energy and the simulated STS pattern for the real Cu phonon spectrum reveal a smoother transition around the Debye energy. Secondly, the simulation assumes a temperature of $T = 0$K. The experimental data is recorded at $T = 6$K, so the experimental data is thermally broadened, which is included in the spectroscopic sections shown in Fig.~\ref{fig-app:EPHThermalBroadening}(c,f).

\clearpage